\documentclass[10pt, journal]{IEEEtran}
\usepackage{amsfonts}
\usepackage{amssymb}
\usepackage[ruled,vlined]{algorithm2e}
\usepackage{mathtools}  
\usepackage{amsthm}
\usepackage[]{algorithm2e}
\usepackage{tabulary}
\usepackage{footnote}
\usepackage[noend]{algpseudocode}
\usepackage{algorithmicx}
\usepackage[export]{adjustbox}
\usepackage{booktabs}
\usepackage{multirow} \usepackage[justification=centering]{caption}
\usepackage{comment}

\usepackage{cases}
\usepackage{cite}
\usepackage{multicol}
\usepackage{xcolor}
\usepackage{graphicx}
\usepackage{tikz}
\usetikzlibrary{arrows.meta,positioning,calc,fit}
\usepackage[caption=false,font=footnotesize]{subfig}
\usepackage{makecell}
\usepackage{bm} 
\usepackage{url}
\graphicspath{{fig/}}

\usepackage{stfloats}
\usepackage{cuted}
\usepackage{lipsum}  
\usepackage{tabularx}

\newcommand{\bele}{\beta_{\rm ele}}
\title{Joint Delay--Doppler Modeling of Multipath LEO Channels
} 

\author{Kuan-Po Chiu, ~\IEEEmembership{Member, ~IEEE}, \; Sumit Roy, ~\IEEEmembership{Fellow, IEEE}
\thanks{Kuan-Po Chiu and Sumit Roy are with the Department of Electrical \& Computer Engineering, University of Washington, Seattle, WA, USA (e-mail:\{kchiu5, sroy@uw.edu\}). Corresponding author: Kuan-Po Chiu.}
}

\begin{document}
\IEEEaftertitletext{\vspace{-0.15in}}
\maketitle
\begin{abstract}
\begingroup
Multipath channel models for Low-Earth-orbit (LEO) non-terrestrial network (NTN) must contend with time variability due to rapid satellite orbital motion that in turn impacts local (near terrestrial) scattering near the ground terminal (GT) receiver. While 3GPP-defined NTN channel model pipelines TR 38.811, 38.901 incorporate elevation-dependence into statistical channel generators, these are
inherited largely from 2-D terrestrial propagation and are largely insufficient. Further, they {\em do not} prescribe any integrated 
workflow that is broadly available for generating {\em joint delay, Doppler/phase and component multipath power} profiles as a function of satellite location over a pass.
This work addresses this gap by developing a geometry-based stochastic channel model driven
by prescribed slant range, elevation, and azimuth over a LEO pass. A ground scene generates wall-reflected, ground-reflected, and
roof-edge-diffracted components, from which joint delay--Doppler statistics
and temporal channel impulse response (CIR) evolution are obtained. The resulting tapped-delay-line CIR representation as a locally  wide-sense stationary tapped-delay-line model, allows exploration of key statistics such as the $K$-factor of the taps, multipath delay and Doppler spreads as a function of elevation and scene parameters.
Depending on the scenarios ({\em urban} and {\em rural}), our model-based results show that with omnidirectional antennas, the LoS probability, delay-spread
and $K$-factor trends broadly match 3GPP NTN models in TR 38.811 as function of elevation. Beamforming instead produces a mid-elevation $K$-factor peak, which is not predicted by TR~38.811~\cite{TR38811} and 6G NTN~\cite{6gntn_d41_2024}.
Doppler spreads are smaller than those from
our previous model~\cite{11534841}, and QuaDriGa~\cite{burkhardt_quadriga_2014} because the proposed model accounts for path existence.  Stationarity durations for CIR are reported and generally decrease with elevation, but with local increases at mid-elevation due to path geometry effects.
\color{black}
\endgroup
\end{abstract}
\vspace*{-0.1in}
\section{Introduction}

LEO downlink channels between a satellite and a GT Rx is fundamentally characterized by a) orbital dynamics and b) terrestrial scattering geometry around the GT location that determine the component multipaths. For a single orbital pass, the time-variations of the resulting stochastic channel model depends on line-of-sight (LoS) direction and the radial and transverse velocity components, that in turn impacts the component path delay and Doppler, that change continuously along the satellite trajectory. 3GPP TR~38.811~\cite{TR38811} seeks to address this by extending the terrestrial TR~38.901~\cite{TR38901} models with environment- and elevation-dependent parameters and an added satellite Doppler shift, allowing a user to generate channel realizations at discrete elevation angles~\cite{TR38811,TR38901}. However, a key gap is that the time correlation statistics for the locally-stationary (elevation-conditioned) realizations is not specified, i.e. how to update channel coefficients as a function of elevation remains open.

Other issues also exist in the 3GPP NTN specification~\cite{TR38811}: some NTN channel parameters in \cite{TR38811} are based 
on ray-tracing results obtained for a 20-km HAPS geometry baseline, which is not applicable to a LEO satellite link~\cite{R1-1803696}. Besides, TR~38.811 assume satellite position is fixed and satellite mobility is omitted~\cite{9815679}. In summary, these multiple limitations contribute to concerns regarding current 3GPP NTN models, in particular around elevation-dependent channel statistics, such as the Rician \(K\)-factor, DS and coherence time.

A popular open-source channel simulator that implements
the 3GPP channel model is QuaDRiGa
\cite{2786139.2786150,burkhardt_mimosadual_2014,burkhardt_quadriga_2014} that has added geometry-based, spatially consistent CIR generation; its NTN extension by Jaeckel \emph{et al.} further adds non-GSO satellite motion, Earth rotation, and satellite orientation~\cite{jaeckel_5g-nr_2022}. Quadriga serves as the natural orbit-aware benchmark for this work, since its  channel parameters are drawn from the 3GPP tables\cite{TR38811}  and inherits every issue noted above.
The proposed model maps a prescribed LoS trajectory and a persistent
terrestrial scene to the power, phase, excess delay, and residual Doppler
of non-direct components. LoS delay and Doppler provide the compensation
reference; the contribution is the resulting multipath channel and its
statistics.

\begin{figure}[h]
    \centering
    \includegraphics[width=0.8\linewidth]{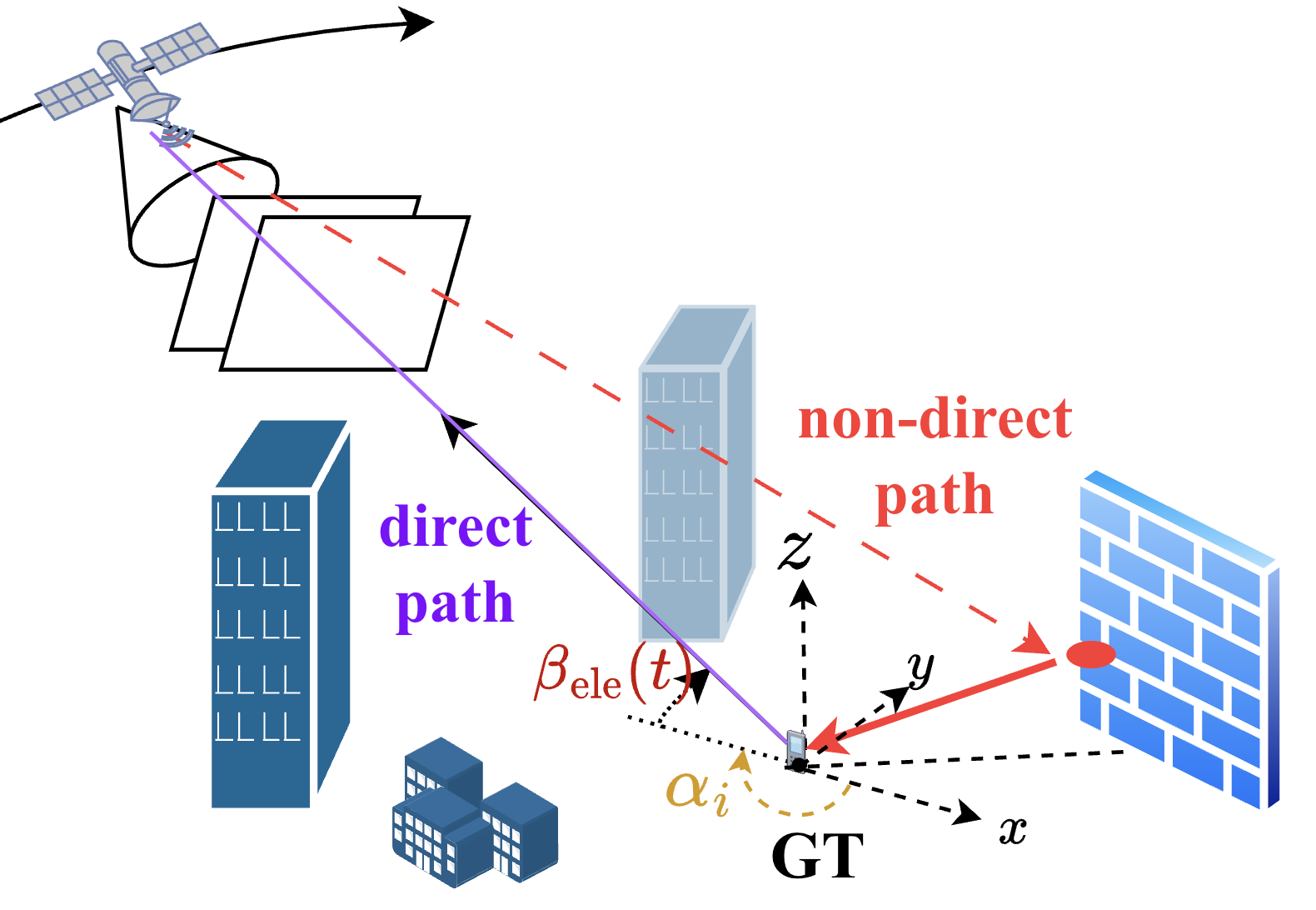}
    \caption{Illustration of non-direct in a scene}
    \label{fig:f1}
\end{figure}
\section{Related Work and Contributions}
\subsection{Prior Work}
\begingroup
Broadly speaking, NTN channel modeling has progressed through a combination of geometry-based stochastic models, analytical formulations, and deterministic ray tracing. We review next several analytical NTN studies that model aspects of NTN channel model without achieving the completeness this work represents.

Ali \emph{et al.} derive orbital Doppler and existence intervals, Khan and Afshang characterize Doppler distributions, and Su \emph{et al.} derive LoS/specular propagation statistics
\cite{ali_doppler_1998,khan_stochastic_2020,su_novel_2025}. Our recent work similarly derives geometry-dependent LEO multipath Doppler properties
\cite{Chiu2025LEO,11534841}. These results capture important marginal or link-level statistics, but not the evolving local multipath CIR of a persistent satellite--ground link.

Closer to the present methodology, Walter \emph{et al.} derive time-varying joint delay--Doppler densities for three-dimensional air-to-air scattering
\cite{walter_joint_2014}, while McBain \emph{et al.} derive delay, Doppler, power-gain statistics, and a scattering function from stochastic mega-constellation geometry
\cite{mcbain_stochastic_2025}. These works establish joint delay--Doppler modeling under mobility, but do not construct a path-persistent terrestrial multipath process driven by a prescribed LoS trajectory with CIR and two-time evolution over one LEO pass.

Ray tracing directly captures such path evolution when a detailed 3-D
environment is available. Ning \emph{et al.} combine orbit propagation and
shooting-and-bouncing rays (SBR) technique to obtain received power, DS, and
multipath Doppler over a transmission window~\cite{10681740,zhang_ray-tracing_2024}.
The 6G-NTN project tabulates urban $K$-factors from ray
tracing~\cite{6gntn_d41_2024}, but its $K$-factor increases monotonically with elevation whereas the TR~38.811 tables~\cite{TR38811} for $K$ peaks at mid elevation.
The conflicting elevation trends reported by these references remain difficult to interpret because the underlying ray-tracing scenes and solver configurations are not sufficiently specified for matched reproduction. The proposed model instead exposes how individual propagation mechanisms contribute to these trends.


\subsection{Contributions}

This paper builds on our earlier
work~\cite{Chiu2025LEO,11534841} to provide a complete NTN modeling approach, in the following ways:
\begin{enumerate}
\item \textbf{New Analytical joint delay--Doppler CIR model:}
We develop a stochastic channel model that maps prescribed
slant range, elevation, and azimuth, together with terrestrial scattering
geometry, to the excess delay, residual Doppler, phase, and power of
multipath components over a LEO pass.
The construction yields a time-varying CIR model, from which estimates of (local) stationarity for joint delay--Doppler statistics are derived. Note that TR~38.811 does not provide any guidance to the time-correlation of the channel representations between these local WSSUS intervals (hence implicitly draws of the channel realizations at successive elevations are independent!) and our work fills this important gap. 

\item \textbf{NTN channel simulation pipeline with open-source code:}
The channel generation code in support of any simulation pipeline is
released as open-source via \footnote{code available: https://github.com/jessest94106/Orbit-Aware-Joint-Delay-Doppler-Channel-Modeling.git. Please check the README for guidance.} with additional README, validation reports and example test scripts, to promote adoption among the community and encourage independent validation of our work. 

\item \textbf{Elevation trends for  NTN channel model parameters:}
Our geometry aware model derivation allows prediction of  the elevation dependence for large-scale channel parameter such
as the $K$-factor and DS, thus offering new insights for interpreting trends reported earlier
by TR~38.811 and other 6G-NTN site-specific ray-tracing studies.

\end{enumerate}
\begin{figure}[!t]
    \centering
    \resizebox{\linewidth}{!}{
\providecommand{\secl}[1]{{\tiny\color{black!60}#1}}
\begin{tikzpicture}[
  font=\scriptsize,
  >={Stealth[length=1.6mm,width=1.4mm]},
  box/.style={draw, rounded corners=1.5pt, align=center, inner sep=3pt,
              text width=#1, fill=white, line width=0.5pt, minimum height=0.9cm},
  core/.style={box=#1, fill=blue!9},
  side/.style={box=#1, fill=gray!8},
  flow/.style={->, line width=0.55pt},
  io/.style={font=\scriptsize, text=red!80!black, inner sep=1.5pt, align=center},
  frame/.style={draw, dashed, rounded corners=3pt, inner sep=5pt, line width=0.5pt},
  flab/.style={font=\scriptsize\itshape, fill=white, inner sep=1.5pt},
]

\node[core=1.85cm] (geo) at (0,0)
  {\textbf{Scatter}\\\textbf{geometry generation}\\[1pt]\secl{Sec.~IV-A--D}};

\node[core=1.65cm, right=1.0cm of geo] (par)
  {\textbf{Path}\\\textbf{parameters}\\[1pt]\secl{Sec.~IV-B--D}};

\node[side=1.65cm, below=0.75cm of geo] (scene)
  {\textbf{Ground scene}\\[1pt]\secl{Sec.~IV-A, drawn once}};

\node[box=1.85cm, below=1.25cm of par] (law)
  {\textbf{Joint delay--}\\\textbf{Doppler law}\\[1pt]\secl{Sec.~V-B, V-C}};

\node[box=1.85cm, below=0.8cm of law] (cir)
  {\textbf{CIR and TDL}\\[1pt]\secl{Sec.~V-A, V-D}};

\node[side=1.75cm, right=1.3cm of cir] (link)
  {\textbf{Model}\\\textbf{evaluation}\\[1pt]\secl{Sec.~VI}};

\draw[flow] ($(geo.west)+(-1.15,0)$) -- (geo.west)
  node[io, pos=0, anchor=south east, yshift=2pt] {$R(t),\;\beta_{\rm ele}(t),\;\alpha_s(t)$};
\draw[flow] (scene) -- (geo)
  node[io, midway, right=1pt] {scatter scene\\(fixed per pass)};
\draw[flow] (geo) -- (par)
  node[io, midway, above=0.85cm] (l2) {Candidate paths,\\interaction points\\blockage};
\draw[flow] (par) -- (law)
  node[io, pos=0.42, right=1pt] {delay, Doppler, phase,\\power per path};
\draw[flow] (law) -- (cir)
  node[io, midway, left=2pt] (l3) {$S_{\mathrm{nd}}(\Delta\tau,\Delta\nu;t)$\\$K(t)$};

\draw[flow] (cir) -- (link)
  node[io, midway, above=1pt] {$h(t,\Delta\tau)$};

\node[frame, fit=(geo)(par)(scene)(law)(cir)(link)(l2)(l3), inner sep=7pt, gray!60] (pass) {};
\node[flab, anchor=south east, text=gray!60!black] at ($(pass.north east)+(0,-0.02)$)
  {one pass: persistent scene, nonstationary channel};
\end{tikzpicture}}
    \caption{ channel workflow driven by $R(t)$, $\bele(t)$, and $\alpha_s(t)$ with a persistent ground scene}
    \label{fig:flow_a}
\end{figure}
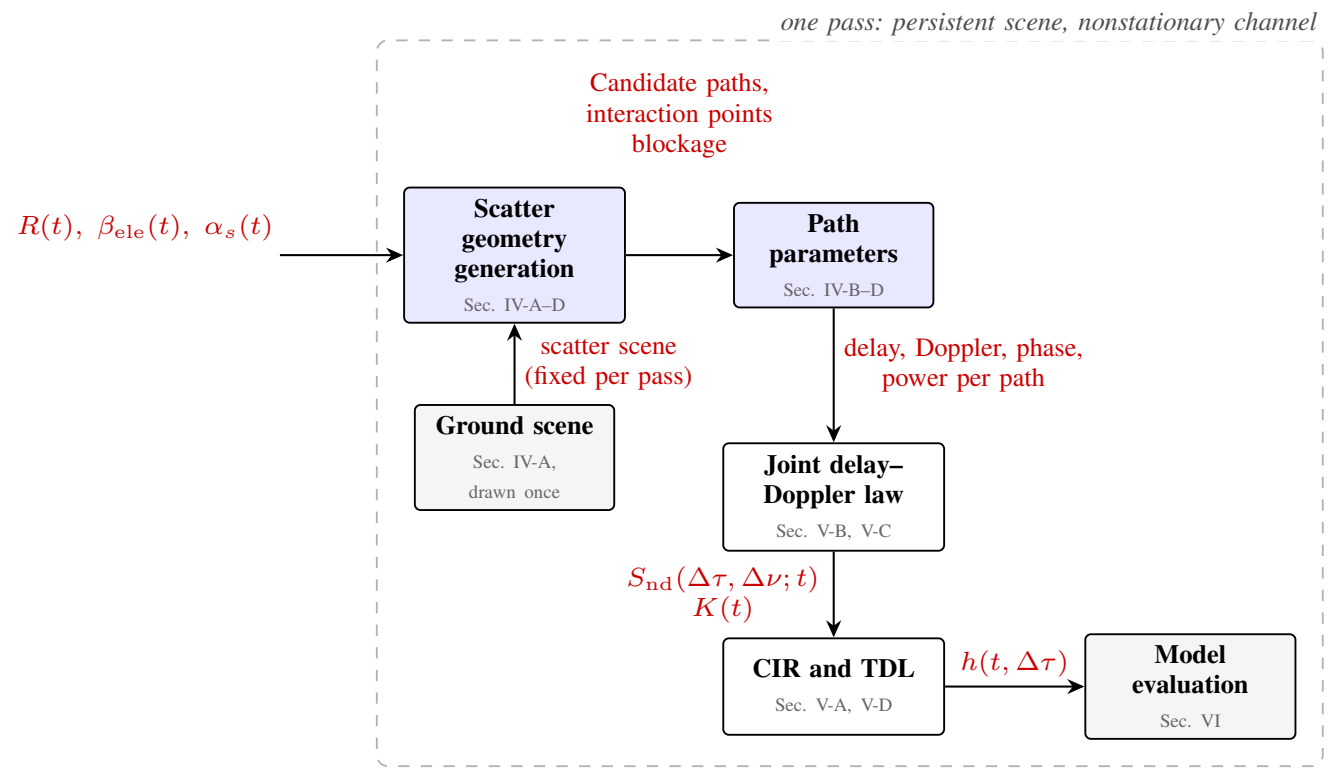

\subsection{Organization of Rest of Paper}
Fig.~\ref{fig:flow_a} summarizes the workflow.
Section~\ref{sec:orbit_dynamics} defines the trajectory inputs and LoS reference;
Section~\ref{sec:persistent_joint_model} constructs the ground scene, path
delay, Doppler, and power, then assembles the CIR and TDL.
Sections~\ref{sec:cir_stats} and~\ref{sec:results} derive the joint
delay--Doppler statistics and evaluate them against reference models.
\endgroup

\section{LoS Doppler}
\label{sec:orbit_dynamics}
\begin{table}[b]
\centering
\caption{Main Notation}
\label{tab:notation}
\footnotesize
\renewcommand{\arraystretch}{1.12}
\begin{tabularx}{\linewidth}{@{}p{0.31\linewidth}X@{}}
\toprule
Symbol & Definition \\
\midrule
$\bele(t),\,\alpha_s(t)$ & Satellite elevation and azimuth; azimuth is measured from $x$ toward $y$. \\
$\mathbf R(t),\,\hat{\mathbf R}(t)$ & LoS displacement vector from the GT antenna to the satellite and its unit vector; the italic $R(t)=\|\mathbf R(t)\|$ is the slant range. \\
$h_g$ & GT antenna height; the ground is at $z=0$ and the antenna is at $(0,0,h_g)$. \\
$\tau_{\rm LoS}(t),\,\nu_{\rm LoS}(t)$ & Direct-path delay and absolute LoS Doppler. \\
$\Phi,\,\lambda_b,\,R_{\rm city}$ & Ground scene, horizontal spatial intensity, and scene radius. \\
$\rho_i,\,\alpha_i,\,\xi_i$ & Wall horizontal distance, location azimuth, and orientation. \\
$H_i,\,\ell_{\rm f},\,\gamma$ & Wall height, width, and Rayleigh height scale. \\
$\mathbf c_i,\,\hat{\mathbf n}_i,\,\hat{\mathbf e}_i$ & Wall base midpoint, wall surface unit normal, and roof-edge unit tangent. \\
$\mathbf p_i^{\rm wall}(t),\,\mathbf p_i^{\rm edge}(t), \beta^b_i$ & Wall-reflection and roof-edge interaction points on wall $i$ and the point elevation relative GT. \\
$d_{s,i}^{b}(t),\,d_{g,i}^{b}(t)$ & Satellite-to-interaction-point and interaction-point-to-GT path lengths, for mechanism $b\in\{\mathrm{wall},\mathrm{edge}\}$. \\
$\Delta\tau_i^b(t),\,\Delta\nu_i^b(t)$ & Excess delay and residual Doppler relative to the LoS reference. \\
\bottomrule
\end{tabularx}
\end{table}

The prescribed slant range $R(t)$, elevation $\bele(t)$, and
azimuth $\alpha_s(t)$ (Fig.~\ref{fig:f1}) can come from SGP4\footnote{SGP4(Simplified General Perturbations~4) predicts satellite
position and velocity from two-line element (TLE) data, accounting
for perturbations such as Earth's oblateness and atmospheric drag.} or the circular-orbit model of
Ali \emph{et al.}~\cite{ali_doppler_1998}. 
Orbit propagation and the detailed LoS Doppler derivation are outside our scope.

The fixed local $x-y-z$ frame has its origin beneath the GT, with ground
$z=0$ and GT position $(0,0,h_g)$.
Azimuth is measured in $x-y$ and elevation from the horizontal;
the GT-to-satellite unit direction is
\begin{equation}
\hat{\mathbf R}(t)=
\begin{bmatrix}
\cos\bele(t)\cos\alpha_s(t)\\
\cos\bele(t)\sin\alpha_s(t)\\
\sin\bele(t)
\end{bmatrix}.
\label{eq:los_unit_vector}
\end{equation}
The LoS vector is $\mathbf R(t)\triangleq R(t)\hat{\mathbf R}(t)$,
placing the satellite at $[0,0,h_g]^{\mathsf T}+\mathbf R(t)$.

A satellite \emph{pass ($\beta_{\rm ele}(t)>0^\circ$) typically last $15$ minutes or less and are usually separated by more than $90$ minutes.}
We neglect Earth rotation within a pass, keep the scene fixed,
and redraw it independently for subsequent pass.


\begingroup
\setlength{\floatsep}{6pt plus 1pt minus 1pt}
\setlength{\textfloatsep}{10pt plus 1pt minus 2pt}
\setlength{\intextsep}{8pt plus 1pt minus 2pt}
\predisplaypenalty=0
\section{Geometry-Based Joint Delay--Doppler Channel Model}
\label{sec:persistent_joint_model}
For carrier frequency $f_c$ and propagation speed $c_0$, the LoS delay and Doppler are
\begin{equation}
\tau_{\rm LoS}(t)=\frac{R(t)}{c_0},\qquad
\nu_{\rm LoS}(t)=-\frac{f_c}{c_0}\dot R(t).
\label{eq:los_doppler_abs}
\end{equation}
\begingroup
Nearby terrestrial objects add non-direct paths due to wall and ground reflections,
and roof-edge diffraction.
With ideal LoS beam tracking at both ends, path $(i,b)$ on wall $i$,
$b\in\{\mathrm{wall},\mathrm{edge}\}$, has excess delay
$\Delta\tau_i^b(t)$ relative to the LoS and residual Doppler
\begin{equation}
\Delta\nu_i^b(t)
\triangleq
\nu_i^b(t)-\nu_{\rm LoS}(t)
=-f_c\frac{d\Delta\tau_i^b(t)}{dt}.
\label{eq:residual_doppler}
\end{equation}

For the ground scene specified in Section~\ref{sec:pdp_ppp}, we next obtain reflection and
diffraction paths in Sections~\ref{sec:specular} and~\ref{sec:edge}), apply the path obstruction in conditions Section~\ref{sec:obstruction} to finally assemble the CIR
in Section~\ref{sec:CIR}.
\endgroup

\subsection{{Ground Scene}}
\label{sec:pdp_ppp}

\begingroup

The built walls in Fig.~\ref{fig:7} are  distributed as 2-D homogeneous spatial Poisson Point Process (PPP) over a circle of radius $R_{\rm city}$ around the GT reference point, i.e. the $i$-th wall base centre $\{\mathbf c_i\}\sim{\rm PPP}(\lambda_b)$ ~\cite{muhammad2017analytical} where $\lambda_b$ is the area intensity representing building density. Wall $i$ is modeled as a vertical rectangle at horizontal range $\rho_i$ and
azimuth $\alpha_i$ from the $x$-axis, with random height $H_i$ and common width $\ell_{\rm f}$. The   scene is thus the spatial process
\begin{equation}
\Phi=\{(\rho_i,\alpha_i,\xi_i,H_i)\colon
0\leq\rho_i\leq R_{\rm city},\ 0\leq\alpha_i<2\pi\}
\label{eq:scene_ppp}
\end{equation}

and orientation $\xi_i\sim{\cal U}[0,\pi)$.
ITU-R P.1410-6~\cite{ituP1410} gives Rayleigh heights with scale $\gamma$:
\begin{equation}
\Pr\{H_i>z\}=\exp\!\left(-\frac{z^2}{2\gamma^2}\right), \; z \,\geq0
\label{eq:itu_height_ccdf}
\end{equation}
Writing $[\cdot]_{xy}$ and $[\cdot]_z$ for horizontal and height components,
the ground-projected center $\mathbf c_i$, with $[\mathbf c_i]_z=0$,
and horizontal unit normal $\hat{\mathbf n}_i$ are
\begin{equation}
\begin{aligned}
\mathbf c_i
&=\rho_i
\begin{bmatrix}
\cos\alpha_i & \sin\alpha_i & 0
\end{bmatrix}^{\mathsf T},\,
\hat{\mathbf n}_i
=
\begin{bmatrix}
-\sin\xi_i & \cos\xi_i & 0
\end{bmatrix}^{\mathsf T}.
\end{aligned}
\label{eq:facade_frame}
\end{equation}
\endgroup

\begin{figure}[t]
    \centering
    \includegraphics[width=\linewidth]{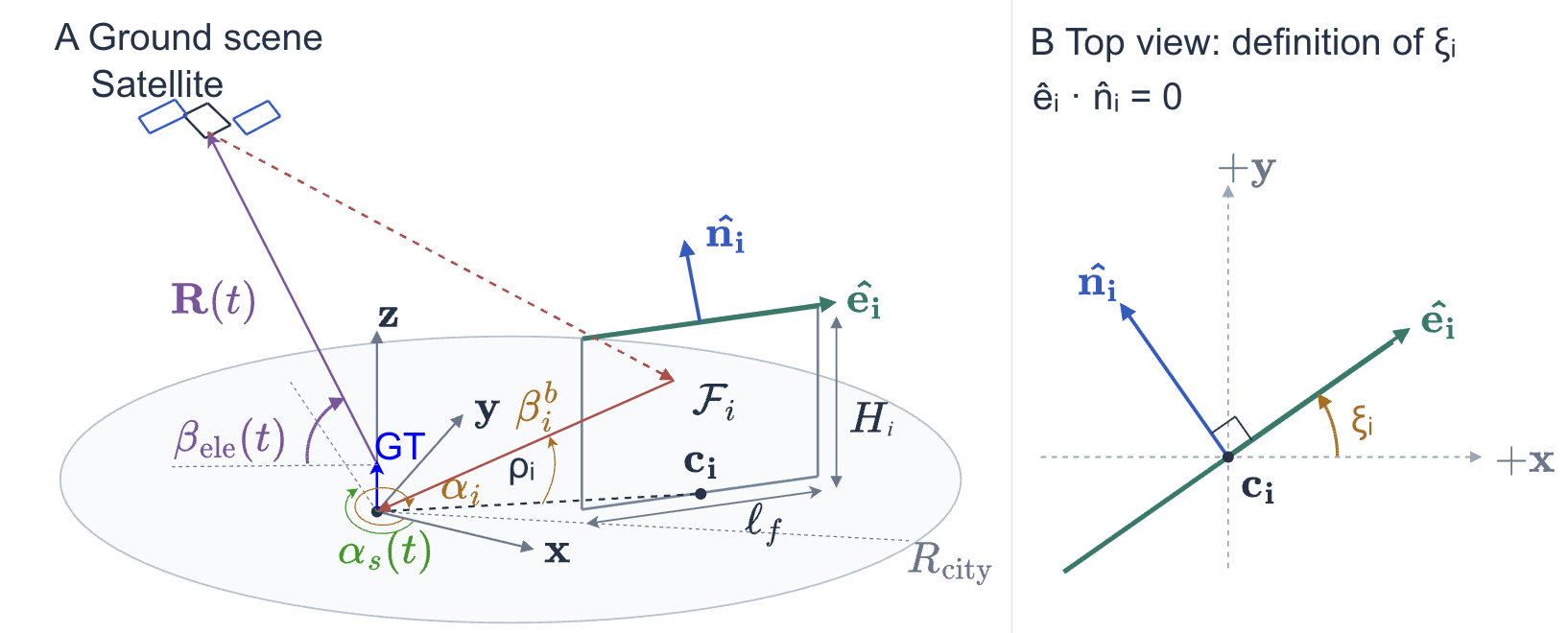}
\caption{Ground-frame scene: wall $i$ base center $\mathbf c_i$ at
$(\rho_i,\alpha_i)$, width $\ell_{\rm f}$, orientation $\xi_i$, and height
$H_i$. The satellite displacement from the GT antenna is
$\mathbf R(t)$.}
    \label{fig:7}
\end{figure}

\begin{figure}[h]
    \centering
    \includegraphics[width=\linewidth]{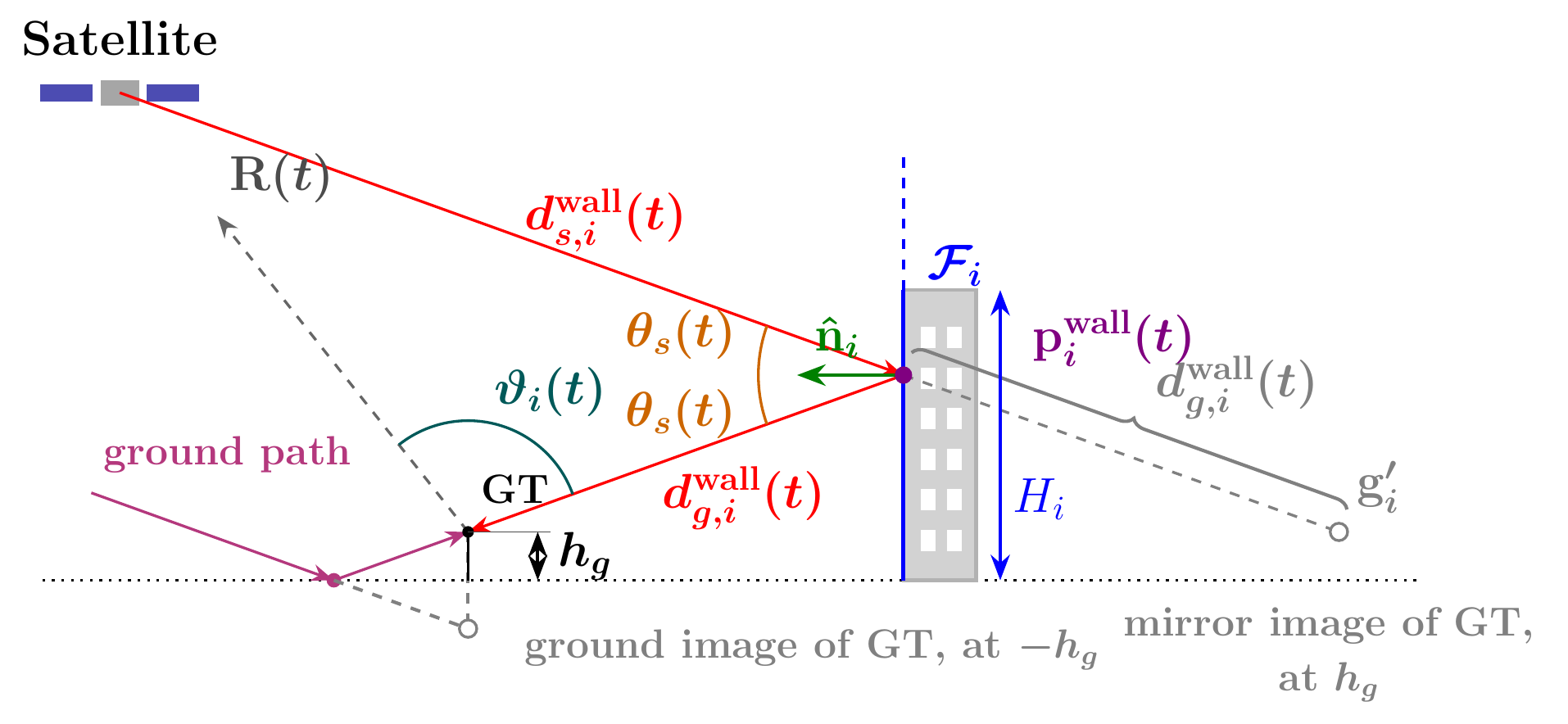}
\caption{The line from
the satellite to the GT image $\mathbf g_i'$ meets the wall plane at the wall-reflection point
$\mathbf p_i^{\rm wall}(t)$.}
\label{fig:mirror}
\end{figure}
Each wall supplies candidate points $\mathbf p_i^{\rm wall}(t)$ and
$\mathbf p_i^{\rm edge}(t)$, for interaction of incipient wavefront.
For $b\in\{\mathrm{wall},\mathrm{edge}\}$, the satellite--interaction-point
and interaction-point--GT path lengths are
{\footnotesize
\begin{equation}
\begin{aligned}
d_{s,i}^b(t)
&=\big|[0,0,h_g]^{\mathsf T}+\mathbf R(t)-\mathbf p_i^b(t)\big|,\,
d_{g,i}^b(t)
=\big|\mathbf p_i^b(t)-[0,0,h_g]^{\mathsf T}\big|,
\end{aligned}
\label{eq:path_unit_directions}
\end{equation}
}
with propagation directions $\hat{\mathbf d}_{s,i}^b(t)
=
\frac{\mathbf p_i^b(t)-[0,0,h_g]^{\mathsf T}-\mathbf R(t)}
{d_{s,i}^b(t)},
\,
\hat{\mathbf d}_{g,i}^b(t)
=
\frac{[0,0,h_g]^{\mathsf T}-\mathbf p_i^b(t)}
{d_{g,i}^b(t)}.$

\subsection{Specular Reflection: Walls and Ground}
\label{sec:specular}

\begingroup

\emph{Wall Reflection.}
A candidate reflected path
must lie on the finite wall rectangle surface ${\cal F}_i$ 
\begin{equation}
\begin{aligned}
&\mathbf p_i^{\rm wall}(t)\in {\cal F}_i=\Big\{\mathbf x:\quad(\mathbf x-\mathbf c_i)^{\mathsf T}\hat{\mathbf n}_i=0,\\
&\quad 0\leq[\mathbf x]_z\leq H_i,\quad\big\|[\mathbf x-\mathbf c_i]_{xy}\big\|
\leq\ell_{\rm f}/2\Big\},
\end{aligned}
\label{eq:facade_rectangle}
\end{equation}
and satisfy the specular direction rule:
\begin{equation}
    \hat{\mathbf d}_{g,i}^{wall}(t)-\hat{\mathbf d}_{s,i}^{wall}(t)
=-2\big[(\hat{\mathbf n}_i)^{\mathsf T}
\hat{\mathbf d}^{wall}_{s,i}(t)\big]\hat{\mathbf n}_i
\label{eq:specular_direction}
\end{equation}
The incoming and outgoing angles equal $\theta_{\rm s}$
(Fig.~\ref{fig:mirror}); walls are large enough for flat-surface Fresnel reflection.

\emph{Delay and Doppler.}
Following appendix~\ref{app:delay_doppler_wall}, the excess delay and residual Doppler are
{\small
\begin{equation}
\begin{aligned}
&\Delta\tau_i^{\rm wall}(t)
=\frac{d_{s,i}^{\rm wall}(t)+d_{g,i}^{\rm wall}(t)-R(t)}{c_0}\\
&\simeq\frac{2\rho_i\,|\sin(\alpha_i-\xi_i)|\,
\cos\bele(t) |\sin(\alpha_s(t)-\xi_i)|}{c_0},\\
&\Delta\nu_i^{\rm wall}(t)
=-f_c\frac{d\Delta\tau_i^{\rm wall}(t)}{dt}\\
&\simeq -f_c\left[-\frac{2\rho_i\sin(\alpha_i-\xi_i)
\cos\bele\sin(\alpha_s (t)-\xi_i)}{c_0}\right]\\
&\times\left[-\dot\bele(t) \frac{\sin\bele(t)}{\cos\bele (t)}
+\dot\alpha_s(t)\frac{\cos(\alpha_s (t)-\xi_i)}{\sin(\alpha_s (t)-\xi_i)}\right]
\end{aligned}
\label{eq:wall_delay_doppler}
\end{equation}}
where the approximation $\rho_i/R(t)\ll1$ has been used.

\emph{Power.}
Far-field rays at the GT and walls are parallel.
With complex relative permittivity
$\widetilde\epsilon_r=\epsilon_r'-j\sigma/(2\pi f_c\epsilon_0)$ and
material parameters from ITU-R P.2040~\cite{ituP2040}, the
polarization-averaged Fresnel power relative to the LoS is
(Appendix~\ref{app:fresnel})
{\small
\begin{equation}
W_i^{\rm wall}(t)=|\Gamma(\theta_{\rm s}(t),\widetilde\epsilon_r)|^2
\left[\frac{R(t)}{d_{s,i}^{\rm wall}(t)+d_{g,i}^{\rm wall}(t)}\right]^2
G\!\left(\vartheta_i(t)\right),
\label{eq:facade_power_weight}
\end{equation}}
where
$|\Gamma(\cdot)|^2=\tfrac12(|\Gamma_{\rm TE}(\cdot)|^2+|\Gamma_{\rm TM}(\cdot)|^2)$,
and $G(\vartheta)$ is the normalized GT array pattern with $G(0)=1$.
\color{black}

\emph{Ground reflection.}
Mirroring the GT from $h_g$ to $-h_g$ gives one ground path with
far-field excess length $2h_g\sin\bele(t)$, so
{\small
\begin{equation}
\begin{aligned}
\Delta\tau^{\rm g}(t)&\simeq\frac{2h_g\sin\bele(t)}{c_0},\\
\Delta\nu^{\rm g}(t)&=-f_c\frac{d\Delta\tau^{\rm g}(t)}{dt}
\simeq-\frac{2f_ch_g\dot\bele(t)\cos\bele(t)}{c_0},
\end{aligned}
\label{eq:ground_record}
\end{equation}}
Appendix~\ref{app:delay_doppler_ground} gives the distance expansion.
The relative power is ($\widetilde\epsilon_r^{\rm g}$ is the ground permittivity)
{\small
\begin{equation}
\begin{aligned}
W^{\rm g}(t)&=|\Gamma(\pi/2-\bele(t),\widetilde\epsilon_r^{\rm g})|^2\\
&\quad\times\left[\frac{R(t)}{R(t)+2h_g\sin\bele(t)}\right]^2
G\!\left(2\bele(t)\right).
\end{aligned}
\label{eq:ground_power}
\end{equation}}
\endgroup

\subsection{Roof-Edge Diffraction}
\label{sec:edge}

\emph{Diffraction point.}
The roof edge has unit direction
$\hat{\mathbf e}_i=[\cos\xi_i\ \ \sin\xi_i\ \ 0]^{\mathsf T}$ (Fig.~\ref{fig:edge_geometry}).
With the propagation directions in \eqref{eq:path_unit_directions},
the diffraction point must lie on the finite edge ${\cal E}_i$
{\small
\begin{equation}
\begin{aligned}
&\mathbf p_i^{\rm edge}(t)\in {\cal E}_i=\Big\{\mathbf x:\quad(\mathbf x-\mathbf c_i)^{\mathsf T}\hat{\mathbf n}_i=0,\\
&\quad [\mathbf x]_z=H_i,\quad\big\|[\mathbf x-\mathbf c_i]_{xy}\big\|
\leq\ell_{\rm f}/2\Big\},
\end{aligned}
\label{eq:edge_cond}
\end{equation}}
 and satisfy the edge-diffraction direction rule:
 {\small
 \begin{equation}
     \hat{\mathbf e}_i^{\mathsf T}
\hat{\mathbf d}_{g,i}^{edge}(t)=\hat{\mathbf e}_i^{\mathsf T}\hat{\mathbf d}_{s,i}^{edge}(t).
\label{eq:29}
 \end{equation}}
At fixed incidence, \eqref{eq:29} holds the outgoing angle to the edge
constant, forming {\em Keller's diffraction cone}~\cite{kouyoumjian_uniform_1974}.

\begin{figure}[h]
    \centering
    \resizebox{0.75\linewidth}{!}{\input{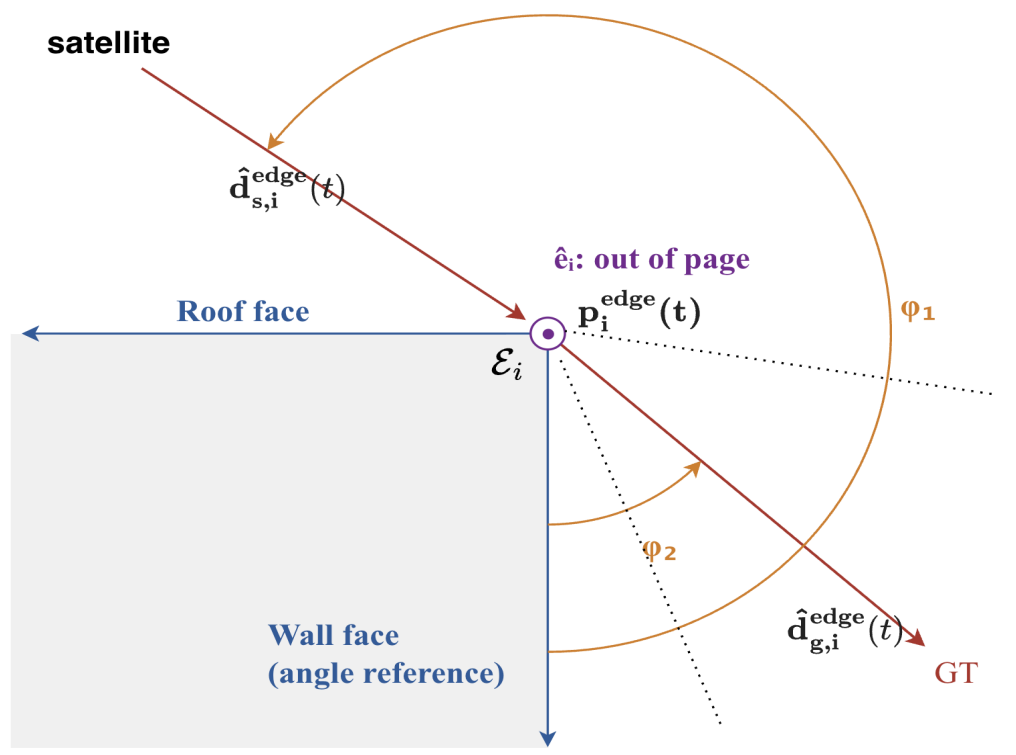}}
    \caption{Roof-edge diffraction geometry projected onto
    the plane normal to the edge. Both angles are measured
    from the wall face through the exterior region.}
    \label{fig:edge_geometry}
\end{figure}

\emph{Delay and Doppler.}
For an admissible edge path, let $\vartheta_i(t)$ be the angle between
the GT-to-edge direction and $\hat{\mathbf R}(t)$. Its excess delay
and residual Doppler are~(Appendix~\ref{app:delay_doppler_edge})
{\footnotesize
\begin{equation}
\begin{aligned}
\Delta\tau_i^{\rm edge}(t)
&=\frac{d_{s,i}^{\rm edge}(t)+d_{g,i}^{\rm edge}(t)-R(t)}{c_0}\simeq\frac{d_{g,i}^{\rm edge}(t)[1-\cos\vartheta_i(t)]}{c_0},\\
\Delta\nu_i^{\rm edge}(t)
&=-f_c\frac{d\Delta\tau_i^{\rm edge}(t)}{dt}\simeq\frac{f_c}{c_0}\dot{\hat{\mathbf R}}{}^{\mathsf T}(t)
\big(\mathbf p_i^{\rm edge}(t)-[0,0,h_g]^{\mathsf T}\big).
\end{aligned}
\label{eq:edge_delay_doppler}
\end{equation}}
LoS-direction evolution has rate
$\|\dot{\hat{\mathbf R}}\|
=\sqrt{\dot\bele^2+\dot\alpha_s^2\cos^2\bele}$.

\emph{Power.}
In Fig.~\ref{fig:edge_geometry}, $\varphi_1$ and $\varphi_2$ measure angles
from the wall face to the projected edge--satellite and edge--GT
directions, respectively; the relative power is
{\small
\begin{equation}
W_i^{\rm edge}(t)=
\frac{\left|D_i\!\left(\varphi_1,\varphi_2,t,f_c\right)\right|^2R^2(t)G\!\left(\vartheta_i(t)\right)}
{d_{s,i}^{edge}(t)d_{g,i}^{edge}(t)
[d_{s,i}^{edge}(t)+d_{g,i}^{edge}(t)]}
.
\label{eq:edge_power_weight}
\end{equation}}
The Uniform Theory of Diffraction (UTD) wedge coefficient
$D_i(\varphi_1,\varphi_2,t,f_c)$~\cite{kouyoumjian_uniform_1974,ituP526}
is specified in Appendix~\ref{app:utd}.

\subsection{Obstruction and Path Existence}
\label{sec:obstruction}

\begin{figure}[!t]
    \centering
    \includegraphics[width=1\linewidth]{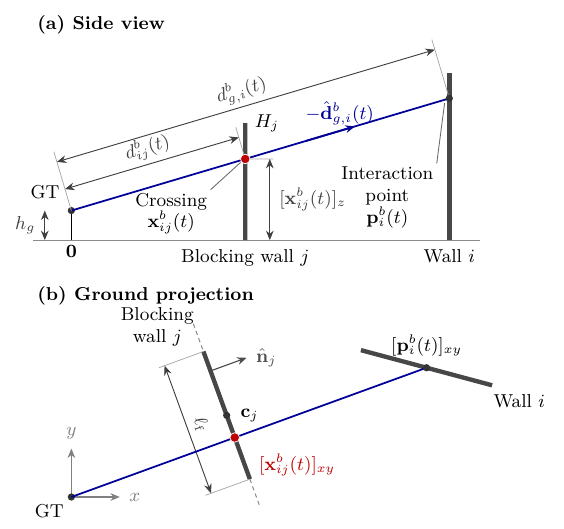}
    \caption{Illustration of Blockage}
    \label{fig:blockage}
\end{figure}

The candidate paths from Sections~\ref{sec:specular} and~\ref{sec:edge}
contribute to the CIR only when path existence conditions are met.
We model blockage only on the interaction-point--GT path, neglecting
satellite--interaction-point blockage.
In Fig.~\ref{fig:blockage}, the GT-to-$\mathbf p_i^b(t)$ ray crosses the
plane of wall $j\ne i$ at $\mathbf x_{ij}^b(t)$, distance $d_{ij}^b(t)$
from the GT along the ray.
Wall $j$ blocks the path if this point lies on the GT--interaction-point
segment and within the wall's width and height:
\color{black}
{\small
\begin{equation}
0<d_{ij}^b<d_{g,i}^b,\;\;
\big\|[\mathbf x_{ij}^b-\mathbf c_j]_{xy}\big\|\leq\ell_{\rm f}/2,\;\;
0\leq[\mathbf x_{ij}^b]_z\leq H_j.
\label{eq:block_edge_down}
\end{equation}}


Define path existence indicator functions as follows: 
{\footnotesize
\setlength{\jot}{0pt}
\setlength{\abovedisplayskip}{3pt plus 3pt minus 1pt}
\setlength{\belowdisplayskip}{3pt plus 3pt minus 1pt}
\setlength{\abovedisplayshortskip}{0pt}
\setlength{\belowdisplayshortskip}{2pt plus 2pt minus 1pt}
\begin{equation}
\begin{aligned}
&\bm1_i^{\rm wall}(t)
=\begin{cases}
1,&\eqref{eq:facade_rectangle},\eqref{eq:specular_direction}\text{ hold, and }\nexists\,j\in\Phi\setminus\{i\}: \eqref{eq:block_edge_down}\text{ holds},\\
0,&\text{otherwise},
\end{cases}\\[1pt]
&\bm1_i^{\rm edge}(t)
=\begin{cases}
1,&\eqref{eq:edge_cond},\eqref{eq:29}\text{ hold, and }\nexists\,j\in\Phi\setminus\{i\}: \eqref{eq:block_edge_down}\text{ holds},\\
0,&\text{otherwise},
\end{cases}\\[1pt]
&\bm1_0(t)=\bm1^{\rm g}(t)
=\begin{cases}
1,&\nexists\,j\in\Phi:\ \eqref{eq:block_edge_down}\text{ holds for the direct LoS ray},\\
0,&\text{otherwise}.
\end{cases}
\end{aligned}
\label{eq:existence}
\end{equation}}
Note that the ground-reflection indicator term $\bm1^{\rm g}(t)$ is assumed approximately equal to direct LoS path existence $\bm{1}_0(t)$.

\subsection{Channel Impulse Response (CIR)}
\label{sec:CIR}

 For one city realization, after compensating the LoS delay and Doppler,
the CIR is
{\footnotesize
\setlength{\abovedisplayskip}{3pt}
\setlength{\belowdisplayskip}{3pt}
\begin{equation}
\begin{aligned}
&h(t,\Delta\tau)
=\underbrace{\bm{1}_0(t)\delta(\Delta\tau)}_{LoS}\\&+\underbrace{\bm{1}_0(t)\sqrt{W^{\rm g}(t)}
e^{j\phi_0^{\rm g}}e^{-j2\pi f_c\Delta\tau^{\rm g}(t)}
\delta\!\left(\Delta\tau-\Delta\tau^{\rm g}(t)\right)}_{h^g \text{(ground)}}\\
&+\underbrace{\sum_{i\in\Phi}\bm{1}_i^{\rm wall}(t)\sqrt{W_i^{\rm wall}(t)}
e^{j\phi_{0,i}^{\rm wall}} e^{-j2\pi f_c\Delta\tau_i^{\rm wall}(t)}
\delta\!\left(\Delta\tau-\Delta\tau_i^{\rm wall}(t)\right)}_{h^{wall}\text{(wall)}}\\
&+\underbrace{\sum_{i\in\Phi}\bm{1}_i^{\rm edge}(t)\sqrt{W_i^{\rm edge}(t)}
e^{j\phi_{0,i}^{\rm edge}} e^{-j2\pi f_c\Delta\tau_i^{\rm edge}(t)}
\delta\!\left(\Delta\tau-\Delta\tau_i^{\rm edge}(t)\right)}_{h^{edge}\text{(edge)}}.
\end{aligned}
\label{eq:hnlos}
\end{equation}}
\color{black}
The powers $W^{\rm g}$ and $W_i^b$ follow \eqref{eq:ground_power},
\eqref{eq:facade_power_weight}, and \eqref{eq:edge_power_weight},
with independent initial phases
$\phi_0^{\rm g},\phi_{0,i}^b\sim\mathcal U[0,2\pi)$.
 For either interaction, the
compensated phase $\phi_i^b(t)=\phi_{0,i}^b-2\pi f_c\Delta\tau_i^b(t)$ satisfies
$\frac{1}{2\pi}\frac{d\phi_i^b(t)}{dt}
=-f_c\frac{d\Delta\tau_i^b(t)}{dt}
=\nu_i^b(t)-\nu_{\rm LoS}(t)=\Delta\nu_i^b(t),$
consistent with \eqref{eq:residual_doppler}. The ground path obeys the
same relation, $\phi^{\rm g}(t)=\phi_0^{\rm g}-2\pi f_c\Delta\tau^{\rm g}(t)$.

The CIR \eqref{eq:hnlos} has two
propagation states: 
{\small
\begin{equation}
\left.h(\cdot)\right|\bm{1}_0=
\begin{cases}
\delta(\Delta\tau)+h^{\rm g}+h^{\rm wall}+h^{\rm edge},
& \bm{1}_0=1~(LoS),\\[4pt]
h^{\rm wall}+h^{\rm edge},
& \bm{1}_0=0~(NLoS).
\end{cases}
\end{equation}}
\color{black}

\subsection{Tapped-Delay-Line (TDL) Representation}
\label{sec:path_tdl}


At discrete times \(t_n=n\Delta t\), where \(n\) is the time index and \(\Delta t\) is the temporal sampling interval, we group the excess delays in \eqref{eq:hnlos} into \(N\) taps of width \(T_s\). Tap \(k\) collects paths with excess delay in \([kT_s,(k+1)T_s)\):
\begin{equation}
h_k[n]
=\int_{kT_s}^{(k+1)T_s}h(t_n,\Delta\tau)\,d\Delta\tau.
\label{eq:tdl_bin}
\end{equation}
Evaluating \eqref{eq:tdl_bin} for the Dirac impulses in \eqref{eq:hnlos}
gives the coherent sum
{\small
\begin{equation}
\begin{aligned}
&h_k[n]={}\bm{1}_0(t_n)\delta_{k0}+\underbrace{\bm{1}_0(t_n)\sqrt{W^{\rm g}(t_n)}
e^{j\phi^{\rm g}(t_n)}
\mathbf{1}\{k^{\rm g}(t_n)=k\}}_{\text{ground}}\\
&+\underbrace{\begin{aligned}[t]
&\sum_{i\in\Phi}\bm{1}_i^{\rm wall}(t_n)\sqrt{W_i^{\rm wall}(t_n)}
e^{j\phi_i^{\rm wall}(t_n)}\mathbf{1}\{k_i^{\rm wall}(t_n)=k\}
\end{aligned}}_{\text{wall}}\\
&+\underbrace{\begin{aligned}[t]
&\sum_{i\in\Phi}\bm{1}_i^{\rm edge}(t_n)\sqrt{W_i^{\rm edge}(t_n)}
e^{j\phi_i^{\rm edge}(t_n)}\mathbf{1}\{k_i^{\rm edge}(t_n)=k\},
\end{aligned}}_{\text{edge}}
\end{aligned}
\label{eq:tdl_cir}
\end{equation}
}
where $k_i^b(t)=\lfloor\Delta\tau_i^b(t)/T_s\rfloor$ and
$k^{\rm g}(t)=\lfloor\Delta\tau^{\rm g}(t)/T_s\rfloor$
index path $(i,b)$ and the ground path; $\delta_{k0}$ places
the compensated LoS in tap $0$.

\color{black}

\section{{Channel Model Statistics}}
\label{sec:cir_stats}

We next derive its first- and
second-order channel statistics for the LoS and NLoS states.

\subsection{Joint Delay--Doppler Power Density}
\label{sec:joint_dd}

From \eqref{eq:hnlos}, 
the scene-averaged non-direct density is
{\footnotesize
\setlength{\jot}{1pt}
\begin{equation}
\begin{aligned}
&S_{\mathrm{nd}}(\Delta\tau,\Delta\nu;t)\\
&\triangleq\mathbb E\!\left[\begin{aligned}
&\underbrace{\bm1_0(t)W^{\rm g}(t)
\delta(\Delta\tau-\Delta\tau^{\rm g}(t))
\delta(\Delta\nu-\Delta\nu^{\rm g}(t))}_{\text{ground}}\\
&+\underbrace{\sum_{i\in\Phi}\bm1_i^{\rm wall}(t)W_i^{\rm wall}(t)
\delta(\Delta\tau-\Delta\tau_i^{\rm wall}(t))
\delta(\Delta\nu-\Delta\nu_i^{\rm wall}(t))}_{\text{wall}}\\
&+\underbrace{\sum_{i\in\Phi}\bm1_i^{\rm edge}(t)W_i^{\rm edge}(t)
\delta(\Delta\tau-\Delta\tau_i^{\rm edge}(t))
\delta(\Delta\nu-\Delta\nu_i^{\rm edge}(t))}_{\text{edge}}
\end{aligned}\right]\\
&=\underbrace{\mathbb E[\bm1_0(t)]W^{\rm g}(t)
\delta(\Delta\tau-\Delta\tau^{\rm g}(t))
\delta(\Delta\nu-\Delta\nu^{\rm g}(t))}_{S^{\rm g}\,\text{(ground)}}\\
&+\underbrace{\begin{aligned}
&\frac{\lambda_b}{\pi}\int_0^\pi\!\int_0^{2\pi}\!\int_0^{R_{\rm city}}
\mathbb E\!\Big[\bm1_i^{\rm wall}(t)W_i^{\rm wall}(t)\\
&\quad\times\delta(\Delta\tau-\Delta\tau_i^{\rm wall}(t))
\delta(\Delta\nu-\Delta\nu_i^{\rm wall}(t))\Big]
\rho_i\,d\rho_i\,d\alpha_i\,d\xi_i
\end{aligned}}_{S^{\rm wall}\,\text{(wall)}}\\
&+\underbrace{\begin{aligned}
&\frac{\lambda_b}{\pi}\int_0^\pi\!\int_0^{2\pi}\!\int_0^{R_{\rm city}}
\mathbb E\!\Big[\bm1_i^{\rm edge}(t)W_i^{\rm edge}(t)\\
&\quad\times\delta(\Delta\tau-\Delta\tau_i^{\rm edge}(t))
\delta(\Delta\nu-\Delta\nu_i^{\rm edge}(t))\Big]
\rho_i\,d\rho_i\,d\alpha_i\,d\xi_i
\end{aligned}}_{S^{\rm edge}\,\text{(edge)}}.
\end{aligned}
\label{eq:joint_measure_definition}
\end{equation}}
Using Campbell--Mecke~\cite{haenggi2012stochastic} to evaluate the expectation of the PPP, at fixed $(\rho_i,\alpha_i,\xi_i)$, whose density follows from \eqref{eq:itu_height_ccdf}, gives

{\footnotesize
\setlength{\jot}{1pt}
\begin{equation}
\begin{aligned}
&S_{\mathrm{nd}}(\Delta\tau,\Delta\nu;t)=\mathbb E[\bm1_0(t)]W^{\rm g}(t)
\delta(\Delta\tau-\Delta\tau^{\rm g}(t))
\delta(\Delta\nu-\Delta\nu^{\rm g}(t))\\
&+\frac{\lambda_b}{\pi}\int_0^\pi\!\int_0^{2\pi}\!\int_0^{R_{\rm city}}
\mathbb E[\bm1_i^{\rm wall}(t)]W_i^{\rm wall}(t)\\
&\quad\times\delta(\Delta\tau-\Delta\tau_i^{\rm wall}(t))
\delta(\Delta\nu-\Delta\nu_i^{\rm wall}(t))
\rho_i\,d\rho_i\,d\alpha_i\,d\xi_i\\
&+\frac{\lambda_b}{\pi}\int_0^\pi\!\int_0^{2\pi}\!\int_0^{R_{\rm city}}\!\int_0^\infty
\frac{H_i}{\gamma^2}e^{-H_i^2/(2\gamma^2)}
\mathbb E[\bm1_i^{\rm edge}(t)\mid H_i]W_i^{\rm edge}(t)\\
&\quad\times\delta(\Delta\tau-\Delta\tau_i^{\rm edge}(t))
\delta(\Delta\nu-\Delta\nu_i^{\rm edge}(t))
\,dH_i\,\rho_i\,d\rho_i\,d\alpha_i\,d\xi_i.
\end{aligned}
\label{eq:joint_measure_averaged}
\end{equation}}
The total expected non-direct power is thus 
\begin{equation}
P_{\mathrm{nd}}(t)=\iint S_{\mathrm{nd}}(\Delta\tau,\Delta\nu;t)
\,d\Delta\tau\,d\Delta\nu.
\label{eq:pnlos}
\end{equation}
Note that in computing Eq~\ref{eq:pnlos} above, the following were used: \\
(i) Ground contribution using the shared direct/ground existence
approximation of Section~\ref{sec:obstruction} yields 
\begin{equation}
\mathbb E[\bm1_0(t)]\simeq
\exp\!\left[-\frac{\sqrt{2/\pi}\lambda_b\ell_{\rm f}\gamma}
{\tan\bele(t)}\operatorname{erfc}\!\left(\frac{h_g}{\sqrt2\gamma}\right)\right].
\label{eq:blockage_B}
\end{equation}
(ii) For an admissible specular reflection from a wall, averaging over its height and
obstruction by other walls gives
{\small
\begin{equation}
\begin{aligned}
&\mathbb E[\bm1_i^{\rm wall}(t)]
\simeq e^{-[\mathbf p_i^{\rm wall}(t)]_z^2/(2\gamma^2)}\\
&\quad\times\exp\!\left[-\frac{\sqrt{2/\pi}\lambda_b\ell_{\rm f}\gamma}{\tan\bele}
\left(\operatorname{erfc}\frac{h_g}{\sqrt2\gamma}\right.\left.-\operatorname{erfc}
\frac{[\mathbf p_i^{\rm wall}(t)]_z}{\sqrt2\gamma}\right)\right].
\end{aligned}
\label{eq:visibility_wall}
\end{equation}}
where the 1st is the probability that a reflection occurs and the second is the probability that no other wall blocks the
reflection-point--GT path segment. See Appendix~\ref{app:visibility} for derivation of the conditional mean of edge factor
$\mathbb E[\bm1_i^{\rm edge}\mid H_i]$.

(iii)Because the ground indicator equals $\bm1_0(t)$ by \eqref{eq:existence} and the wall- and edge-path statistics are only weakly dependent on the direct-path state in the considered geometry; we therefore approximate their conditional densities:
{\small
\begin{equation}
\begin{aligned}
S_{\mathrm{nd}}&\approx\mathbb E[\bm1_0]\,S_{\mathrm{nd}}^{\rm L}+\big(1-\mathbb E[\bm1_0]\big)\,S_{\mathrm{nd}}^{\rm N},  {\rm where}\\
S_{\mathrm{nd}}^{\rm L}&=W^{\rm g}(t)\,\delta(\Delta\tau-\Delta\tau^{\rm g}(t))\delta(\Delta\nu-\Delta\nu^{\rm g})
+S^{\rm wall}+S^{\rm edge},\\
S_{\mathrm{nd}}^{\rm N}&=S^{\rm wall}+S^{\rm edge},
\end{aligned}
\label{eq:two_state}
\end{equation}}
where $S_{\mathrm{nd}}^{\rm L}$ and $S_{\mathrm{nd}}^{\rm N}$ are the non-direct densities
given a LoS and NLoS, with powers $P_{\mathrm{nd}}^{\rm L}$ and
$P_{\mathrm{nd}}^{\rm N}$, respectively. 

\subsection{K-factor, Delay Spread, Doppler Spread, and Coherence Time}
\label{sec:power_stats}

With unit direct-path power, the K-factor is the direct-path power over the
non-direct power given a clear direct path, the convention of~\cite{TR38811},
\begin{equation}
K(t)=\frac{1}{P_{\mathrm{nd}}^{\rm L}(t)}.
\label{eq:K_factor}
\end{equation}

For $q\in\{\mathrm L,\mathrm N\}$ denoting the LoS and NLoS states,
let $P_{\mathrm{nd}}^q(t)=\iint S_{\mathrm{nd}}^q(\Delta\tau,\Delta\nu;t)
\,d\Delta\tau\,d\Delta\nu$.
The normalized non-direct delay profile in each state is
$\widetilde P_\tau^q(\Delta\tau;t)=
\int S_{\mathrm{nd}}^q(\Delta\tau,\Delta\nu;t)\,d\Delta\nu/P_{\mathrm{nd}}^q(t)$.
With total powers $P^{\rm L}(t)=1+P_{\mathrm{nd}}^{\rm L}(t)$ and
$P^{\rm N}(t)=P_{\mathrm{nd}}^{\rm N}(t)$, the mean excess delay
and RMS delay spread of the full channel in each state are
{\small
\begin{equation}
\begin{aligned}
\bar\tau^q(t)
&=\frac1{P^q(t)}\iint \Delta\tau\,S_{\mathrm{nd}}^q(\Delta\tau,\Delta\nu;t)
\,d\Delta\tau\,d\Delta\nu,\\
[\mathrm{DS}_{\mathrm{RMS}}^q(t)]^2
&=\frac1{P^q(t)}\iint \Delta\tau^2S_{\mathrm{nd}}^q(\Delta\tau,\Delta\nu;t)\\
&\qquad\times d\Delta\tau\,d\Delta\nu-[\bar\tau^q(t)]^2.
\end{aligned}
\label{eq:rms_width}
\end{equation}}

Define the normalized non-direct Doppler marginal in each state as
$\widetilde P_\nu^q(\Delta\nu;t)=
\int S_{\mathrm{nd}}^q(\Delta\tau,\Delta\nu;t)\,d\Delta\tau/
P_{\mathrm{nd}}^q(t)$ and its mean as
$\bar\nu^q(t)=\int\Delta\nu\,\widetilde P_\nu^q(\Delta\nu;t)\,d\Delta\nu$.
The non-direct RMS Doppler spread and inverse-spread coherence scale are
\begin{equation}
\begin{aligned}
[\sigma_\nu^q(t)]^2
&=\int[\Delta\nu-\bar\nu^q(t)]^2
\widetilde P_\nu^q(\Delta\nu;t)\,d\Delta\nu,\\
T_c^q(t)&=\frac{1}{2\pi\sigma_\nu^q(t)}.
\end{aligned}
\label{eq:coherence_rms_scale}
\end{equation}

\subsection{CIR Stationarity Interval}
\label{sec:stationarity}
While the coherence scale in \eqref{eq:coherence_rms_scale}
characterizes how quickly channel coefficients decorrelate due to Doppler spread, the CIR in \eqref{eq:hnlos} also evolves due to
changes in path geometry/availability as a function of elevation. Hence, persistence of CIR delay--Doppler is captured via the stationarity interval for the non-direct paths as follows.

To select a local WSSUS interval, we evaluate the
normalized correlation between non-direct
delay--Doppler power density at two times,
inspired by~\cite{bernado_wssus_2012,10225614}.
For each scene $\Phi$, we apply Gaussian smoothing:
\begin{equation}
C_t=\mathcal G*S_{\mathrm{nd}}(\cdot;t\mid\Phi),
\end{equation}
where $S_{\mathrm{nd}}(\cdot;t\mid\Phi)$ is obtained
from the non-direct terms of $h$ in \eqref{eq:hnlos},
and $*$ denotes convolution over delay and Doppler.
The Gaussian kernel
$\mathcal G(\tau,\nu)=
\frac{1}{2\pi T_s\sigma_\nu}
\exp[-\tau^2/(2T_s^2)-\nu^2/(2\sigma_\nu^2)]$
smooths each scene's delay--Doppler power profile
to represent finite delay and Doppler resolution.
After an elapsed time $u$, the normalized similarity,
averaged over scenes, is
\begin{equation}
F(t,u)=
\mathbb E_{\Phi}\!\left[
\frac{\langle C_t,C_{t+u}\rangle}
{\|C_t\|_2\,\|C_{t+u}\|_2}
\right].
\label{eq:migrated_fraction}
\end{equation}
Here, $F(t,u)=1$ indicates full correlation between
the normalized power density. The stationarity
interval is the first elapsed time at which this
correlation falls to or below a prescribed threshold
$\epsilon\in(0,1)$:
\begin{equation}
T_{\rm sta}(t)=
\inf\{u>0:F(t,u)\leq\epsilon\}.
\label{eq:stationarity_time}
\end{equation}

\endgroup
 \section{Model Validation and Numerical Results}
\label{sec:results}

\begingroup

Unless stated otherwise, all results use an overhead pass ( $\alpha_s(t)=0$) with
$h_s=550$~km, $h_g=1.5$~m, $f_c=2$~GHz, $R_{\rm city}=2$~km,
and an omnidirectional GT antenna.
The urban scene uses $\gamma=8$~m~\cite{4483593},
$\lambda_b=1.96\times10^{-3}$~m$^{-2}$, and $\ell_{\rm f}=20$~m;
the rural scene uses $\gamma=3$~m, $\lambda_b=3.14\times10^{-3}$~m$^{-2}$, and $\ell_{\rm f}=10$~m. Their permittivity follows P.2040~\cite{ituP2040}; the ground relative permittivity is $3.5$ for both scenes.
Table~\ref{tab:model_elevation_comparison} summarizes the resulting trends as a function of elevation.
\endgroup

\begin{table*}[!t]
\centering
\caption{Trends in channel statistics as $\bele$ increases.}
\label{tab:model_elevation_comparison}
\footnotesize
\setlength{\tabcolsep}{3pt}
\renewcommand{\arraystretch}{1.05}
\begin{tabularx}{\textwidth}{@{}
  >{\raggedright\arraybackslash}p{0.09\textwidth}
  >{\raggedright\arraybackslash}p{0.19\textwidth}
  >{\raggedright\arraybackslash}p{0.13\textwidth}
  >{\raggedright\arraybackslash}p{0.09\textwidth}
  >{\raggedright\arraybackslash}p{0.13\textwidth}
  >{\raggedright\arraybackslash}X@{}}
\toprule
\textbf{Metric} & \textbf{Proposed model}
& \textbf{3GPP TR 38.811~\cite{TR38811}}
& \textbf{6G-NTN}\par\cite{6gntn_d41_2024}
& \textbf{QuaDRiGa~\cite{jaeckel_5g-nr_2022}}
& \textbf{Comments} \\
\midrule
LoS/NLoS probability
& Increase in LoS; rural LoS probability is higher than urban.
& Increase in LoS; suburban/rural LoS probability is higher than urban.
& ---
& Reuses TR~38.811 LoS/NLoS probabilities.
& Proposed rural closely matches TR~38.811; urban trends agree with TR~38.811 and 6G-NTN.
Magnitude differences may reflect scene settings. \\
\addlinespace[3pt]
$K$-factor
& LoS omni: increases in rural/urban scenes.\par Beamforming urban: increases to a maximum near $40^\circ$, then decreases.
& Urban/rural: decreases.\par Dense urban: increases up to $30^\circ$, then decreases.
& Urban: generally increases~\cite[Fig.~44]{6gntn_d41_2024}. \par
Rural:---
& Decrease in rural/urban.
& Proposed model with omni agrees with 6G-NTN and TR~38.811 dense urban but differs from TR~38.811 urban/rural.
QuaDRiGa decreases because its LoS-state non-direct power is incorrectly derived from the TR~38.811 NLoS clutter loss~\cite{jaeckel_5g-nr_2022}; \\
\addlinespace[3pt]
RMS delay spread
& LoS: urban decreases from $45$ to $3.7$~ns; rural decreases from $13$ to about $4$~ns.
\par NLoS: decreases up to $70^\circ$ in urban and $40^\circ$ in rural, then flatten.
& In LoS/NLoS states, urban generally decreases and rural generally increases.
& Urban: decreases~\cite[Fig.~45]{6gntn_d41_2024}.\par
Rural:---

& LoS: Rural/urban increase; rural much larger. NLoS: ---. \par PDP: no elevation dependence.
& Proposed urban LoS agrees with TR~38.811 and 6G-NTN; rural differs from TR~38.811.
QuaDRiGa's PDP misses elevation dependency. \\
\addlinespace[3pt]
RMS Doppler spread / $T_c$
& LoS: urban spread increases up to $50^\circ$, then decreases; rural increases.
\par NLoS: urban increases and flattens at high elevation; rural increases generally.
& Jakes at fixed GT speed~\cite{TR38901}, plus common satellite Doppler shift. No satellite-induced Doppler spread.
& ---
& LoS: urban fluctuates weakly; rural: increases.
& Proposed model captures the geometry dependency of Doppler spread while \cite{Chiu2025LEO} only gives the upper bound.
QuaDRiGa gives $1$--$3$ orders larger LoS-state spreads that align with the analytical upper bound model in \cite{Chiu2025LEO} \\
\addlinespace[3pt]
Stationarity time $T_{\rm sta}$
& Urban: generally decreases, with a temporary increase near
$60^\circ$.
\par Rural: generally decreases, with a temporary increase near
$50^\circ$--$55^\circ$.
& --- & --- & ---
& We characterize the elevation dependence of the proposed joint delay–Doppler path-persistence time to guide local WSSUS window selection. \\
\bottomrule
\end{tabularx}
\par\vspace{2pt}
\begin{minipage}{\textwidth}
\scriptsize\raggedright
 The 6G-NTN urban curves use $3.5$~GHz,
isotropic antennas, and ray tracing with diffuse scattering~\cite[Sec.~4.2]{6gntn_d41_2024}.
Stationarity is unconditional, with $\epsilon=0.99$, $T_s=33$~ns, and $NT_s=10$~s.
``---'' denotes not reported.
\end{minipage}
\end{table*}

\subsection{Joint Delay--Doppler Power Spectral Density}
\label{sec:res_joint}

\begingroup

Fig.~\ref{fig:joint_dd} resolves the three non-direct contributions under LoS, i.e. Eq.
\eqref{eq:joint_measure_definition} - for $20^\circ$ and $80^\circ$ elevation.
Wall-reflected power concentrates near zero excess delay as elevation increases (the excess delay below which $95\%$ of the wall-reflected power lies decreases from $170$ to $2.0$~ns) naturally because increasing elevation reduces the excess delay of wall-reflected paths, as shown in \eqref{eq:wall_delay_doppler}. Moreover, vertical-wall reflection relative Doppler is proportional to excess delay when $\dot\alpha_s=0$ ~\eqref{eq:wall_delay_doppler} (overhead pass orbit)\footnote{Note that for a non-overhead pass with $\dot\alpha_s\neq0$, the wall-orientation-dependent term in \eqref{eq:wall_delay_doppler} introduces Doppler spread at a given excess delay, while ground reflection Doppler shift remains invariant.}, i.e. straight line with slope determined by the elevation.

In contrast, roof-edge diffraction is weaker but spans a wider Doppler range at a given excess delay. The diffraction condition in \eqref{eq:29} constrains only the propagation-direction component along the roof edge, allowing different arrival elevations. Consequently, Doppler is no longer fixed by delay alone in \eqref{eq:edge_delay_doppler}.

Ground reflection contributes single Doppler value through \eqref{eq:ground_record} for all excess delays. Its excess path length $2h_g\sin\bele$ increases with elevation, from $1.03$~m at $20^\circ$ to $2.95$~m at $80^\circ$, corresponding to excess delays of $3.42$ and $9.85$~ns, respectively.

\begin{figure}[h!]
    \centering
    \includegraphics[width=\linewidth]{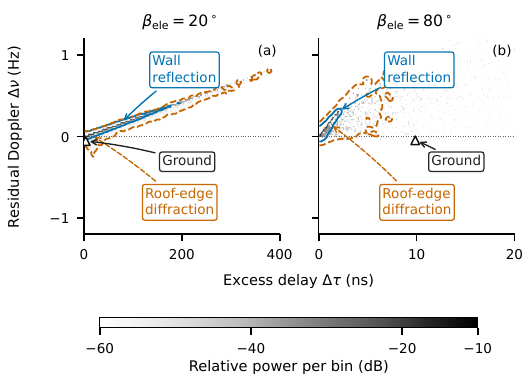}
\caption{Joint non-direct power \eqref{eq:joint_measure_definition} in the LoS state, overhead pass.
Grey: power per bin ($2.5$~ns at $20^\circ$, $0.125$~ns at $80^\circ$, by $0.02$~Hz), normalized by $P_{\mathrm{nd}}^{\rm L}$. Blue/orange:
$90\%$ wall-reflection/roof-edge contours; triangles: ground reflection.}
    \label{fig:joint_dd}
\end{figure}

\endgroup
\subsection{Non-Direct Component Power and Path-Count Contribution}
\label{sec:res_share}

\begin{figure}[!t]
    \centering
    \includegraphics[width=\linewidth]{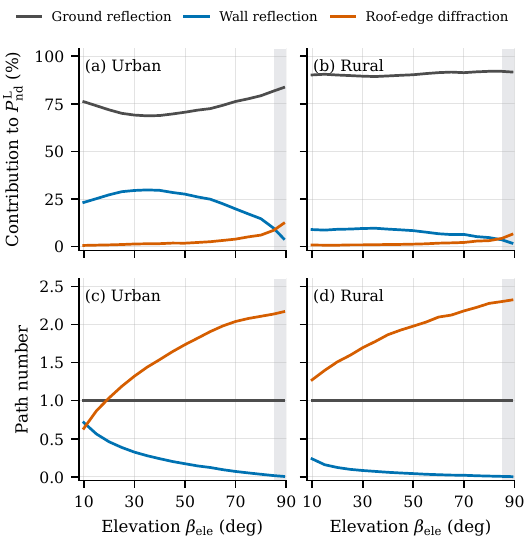}
    \caption{LoS-state non-direct contributions: (a), (b) power shares;
    (c), (d) expected valid-path count shares.}
    \label{fig:power_path}
\end{figure}

\begingroup

Fig.~\ref{fig:power_path} separates power and path-count contributions in
the LoS state of \eqref{eq:two_state}. Power shares are normalized by
$P_{\mathrm{nd}}^{\rm L}$; count shares use the expected number of valid
paths from \eqref{eq:existence}, including one ground path.

In urban cases, ground reflection carries $69$--$79\%$ of the non-direct power
between $\beta_{ele}(t)=10^\circ/80^\circ$. Wall reflection rises from $23\%$ to
about $30\%$ near $\beta_{ele}(t)=35^\circ$, then falls to $15\%$ as the blockage in \eqref{eq:visibility_wall} suppresses long paths. In contrast, roof-edge diffraction contributes about $6\%$ of the non-direct power at $\beta_{\rm ele}(t)=80^\circ$, with arrivals extending beyond the delay range of wall reflection in Fig.~\ref{fig:joint_dd}. 

In the rural scene, lower wall heights reduce the probability of supporting a specular reflection through \eqref{eq:visibility_wall}, while roof-edge diffraction remains weak. Ground reflection therefore dominates, contributing $89$--$92\%$ of the non-direct power over the evaluated elevation range.

Expectations of path counts provide different insights.
Roof-edge diffraction generally produces more paths than wall reflection in both urban and rural scenes, providing more opportunities for multipath dispersion. Wall reflection is most likely at low elevations and is more prevalent in the urban scene, where taller buildings allow more valid specular reflection points.
\color{black}

Summary: small expected counts indicate limited multipath in LEO downlink for the adopted geometry and visibility model, despite the large downlink beam footprint.
\color{black}

\endgroup

\subsection{LoS and NLoS State Probabilities}
\label{sec:res_plos}

\begin{figure}[!t]
    \centering
    \includegraphics[width=\linewidth]{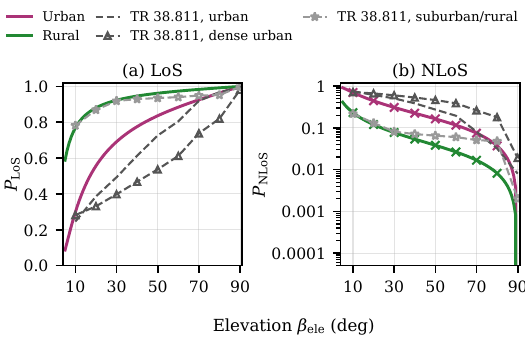}
    \caption{State probabilities from \eqref{eq:blockage_B}: (a) $P_{\rm LoS}=\mathbb E[\bm1_0]$ and
    (b) $P_{\rm NLoS}=1-\mathbb E[\bm1_0]$ against the TR~38.811 Table~6.6.1-1. All continuous lines are analytical model}
    \label{fig:plos}
\end{figure}

\begingroup

Fig.~\ref{fig:plos} compares the urban and rural LoS/NLoS probabilities versus elevation. In both environments, higher elevation increases the LoS probability due to lower blockage.
Compared to urban, the shorter and narrower walls in rural yield a smaller blockage factor in \eqref{eq:blockage_B}, giving a higher LoS probability. The urban--rural difference narrows as both LoS probabilities approach 1 near $\bele=90^\circ$.
\color{black}

Here the rural LoS probabilities closely follow the TR~38.811 suburban/rural values while the urban curve captures the increasing LoS and decreasing NLoS trends~\cite{TR38811}. 
Our results thus provide causative explanations for the trends as a function of building-height distribution, wall widths, and spatial density.
\endgroup

\subsection{ Channel Impulse Response and TDL Realization}
\label{sec:res_cir}

\begin{figure}[h]
    \centering
    \includegraphics[width=1\linewidth]{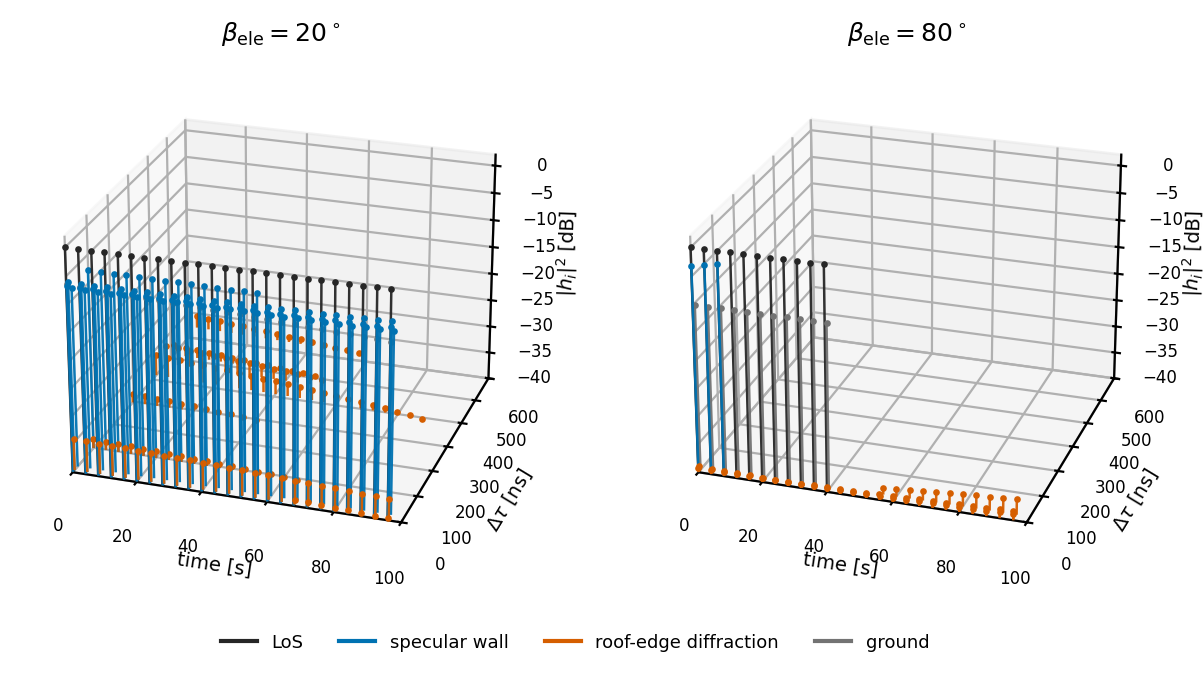}
\caption{Pathwise CIR over $100$~s in one fixed scene, initially at
$20^\circ$ (left) and $80^\circ$ (right). Black: direct; blue: wall;
orange: roof edge; grey: ground.}
    \label{fig:cir}
\end{figure}

\begingroup

Fig.~\ref{fig:cir} shows a CIR realization of \eqref{eq:hnlos}, separating the direct and non-direct components at $\beta_{ele}(t)=20^\circ$ and $80^\circ$. Ground and wall reflections produce stronger power at shorter excess delays, while roof-edge diffraction produces weaker power over a wider delay range, which is consistent with Fig.~\ref{fig:power_path}.

The blockage effect can be seen in the $\beta_{ele}(t)=80^\circ$ plot.\color{black} As the satellite moves, the wall-reflected paths disappear within the first few seconds of excess delay. Near $t=40$~s, direct-path blockage removes both the direct and ground terms, leaving only roof-edge paths in this realization.
\endgroup

\begin{figure}[t]
    \centering
    \includegraphics[width=1\linewidth]{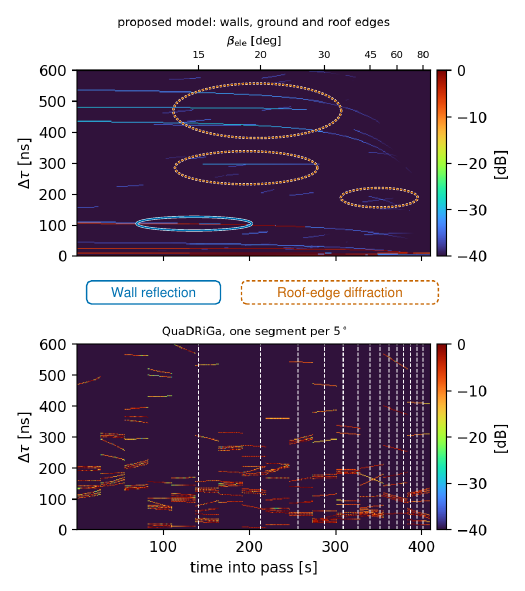}
\caption{TDL evolution over a pass: proposed scene model (top) and
QuaDRiGa NTN urban LoS with independent $5^\circ$ elevation draws
(bottom; dashed boundaries).}
    \label{fig:waterfall}
\end{figure}

\begingroup

Fig.~\ref{fig:waterfall} shows TDL realization over a satellite pass.
The path existence and delays evolve continuously as per 
\eqref{eq:hnlos} and impacts the {\em local stationarity intervals discussed further in Sec. I}. Note that other common NTN CIR generation tools such as  QuaDRiGa~\cite{jaeckel_5g-nr_2022} only generates {\em independent} channel realizations per elevation \footnote{Although QuaDRiGa can evolve the CIR over a full pass in one
segment, it draws the initial large-scale parameters from the tables
only once, at the beginning of the segment. Thus, statistics such as
delay spread may not evolve correctly with elevation.} (represented by the dashed vertical line boundaries in Fig.~\ref{fig:waterfall}) using TR~38.811 Table \cite{TR38811}, in effect ignoring channel stationarity duration. 

\endgroup

\subsection{Power Delay Profile versus Elevation}
\label{sec:res_pdp}

\begin{figure}[t]
    \centering
    \includegraphics[width=\linewidth]{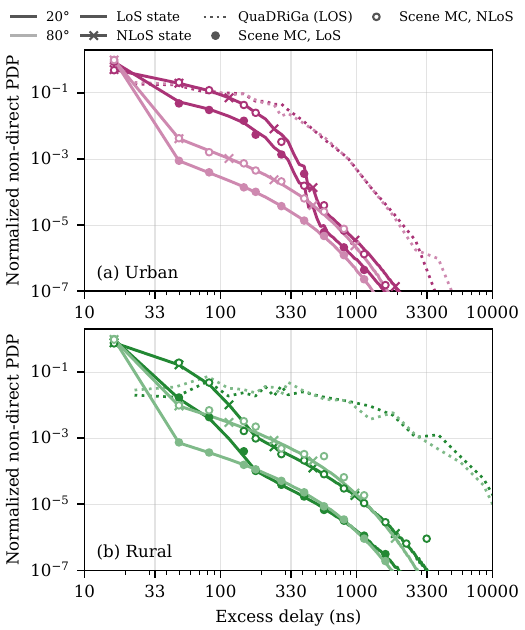}
    \caption{Normalized non-direct PDPs in (a) urban and (b) rural scenes at
    $20^\circ$ (Darker color) and $80^\circ$ (Lighter color),  $33$~ns per bin. Continuous lines: analytical
    model LoS state; Continuous lines with X: analytical model NLoS state.  Circle: scene simulation; dotted: QuaDRiGa NTN LoS~\cite{jaeckel_5g-nr_2022}. Each curve is normalized by its
    non-direct power.}
    \label{fig:pdp_elevation}
\end{figure}

\begingroup

Fig.~\ref{fig:pdp_elevation} shows the normalized non-direct PDPs in the LoS and NLoS states, where only the LoS state includes ground reflection. As elevation increases, the urban PDP becomes more concentrated at smaller delays as wall-reflection delays decrease. In rural case, the shorter walls further limit large delay from wall reflections.

In the LoS state, the ground delay in \eqref{eq:ground_record} is below $10$~ns, so the ground reflection falls in the first bin. Its absence in the NLoS state reduces the first-bin power, giving the wall and edge contributions a larger share of the normalized PDP.

The circles represents results from MC simulations based on \eqref{eq:scene_ppp}, the path conditions, and blockage \eqref{eq:existence}. With $20000$ realizations per elevation for each environment, the simulated PDPs closely follow the analytical profiles.
In contrast, the QuaDRiGa PDP extends to longer delays in rural than in urban and shows weak elevation dependence in both scenes~\cite{jaeckel_5g-nr_2022}.
\endgroup

\subsection{Delay Spread versus Elevation}

\begin{figure}[!h]
    \centering
    \includegraphics[width=\linewidth]{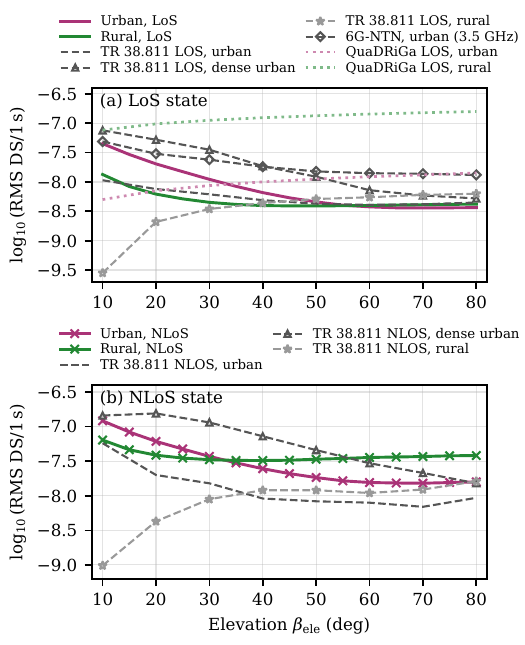}
    \caption{RMS DS \eqref{eq:rms_width} in (a) LoS and (b) NLoS states.
    References: corresponding TR~38.811 S-band rows, QuaDRiGa LOS and urban 6G-NTN at $3.5$~GHz.}
    \label{fig:delay_spread}
\end{figure}

\begingroup

Fig.~\ref{fig:delay_spread} compares the LoS- and NLoS-state RMS
delay spreads using \eqref{eq:rms_width}. As elevation increases, both urban and rural LoS-state DS decrease and then remain nearly constant after $40^\circ$ elevation. Besides, LoS-state
DS is smaller because the ground paths concentrate
power at short delays. Without it, NLoS-state
DS reflects the broader spread from wall/edge. 

The proposed urban LoS- and NLoS-state DS broadly follow the
decreasing trends of the corresponding TR~38.811
references~\cite{TR38811} and, for LoS,
6G-NTN~\cite{6gntn_d41_2024} as elevation increases. The proposed rural LoS-state DS,
however, decreases while TR~38.811 predicts an increase.
In contrast, QuaDRiGa predicts
increasing DS in both scenes~\cite{jaeckel_5g-nr_2022}
but exhibits an order-of-magnitude mismatch to ~\cite{TR38811} and ~\cite{6gntn_d41_2024}.

Since the scene parameters are not disclosed in
TR~38.811~\cite{TR38811} and 6G-NTN~\cite{6gntn_d41_2024},
the results assess agreement in trends and magnitudes rather
than validate the model under matched conditions.

\endgroup

\subsection{K-factor versus Elevation}
\label{sec:res_K}

\begin{figure}[!h]
    \centering
    \includegraphics[width=\linewidth]{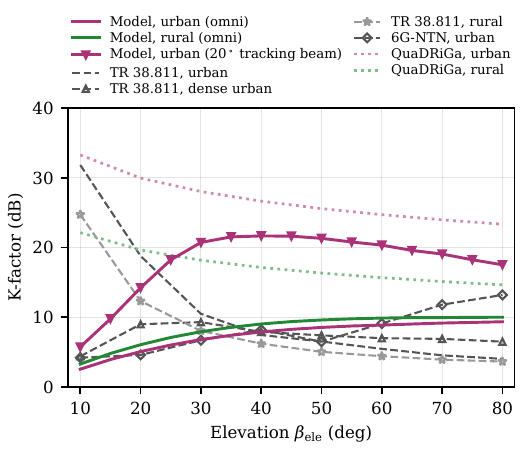}
    \caption{K-factor \eqref{eq:K_factor} in the LoS state, with unit direct-path reference
    power. The urban beamforming case uses $\sigma=20^\circ$; the remaining model curves are omni.}
    \label{fig:k_validation}
\end{figure}

\begingroup
Fig.~\ref{fig:k_validation} shows the LoS-state K-factor.
This follows the trends of 
6G-NTN trend and TR~38.811 dense urban, whereas the TR~38.811 urban and QuaDRiGa curves decrease
\cite{TR38811,6gntn_d41_2024,jaeckel_5g-nr_2022}.

The urban beamforming case uses the normalized power pattern
$G(\vartheta)=e^{-\vartheta^2/(2\sigma^2)}$ with
$\sigma=20^\circ$, steered along the LoS direction.
At fixed elevation, narrowing the beam suppresses off-LoS power
and increases $K$, while widening it approaches the omnidirectional
result. Thus, beamforming raises $K$ above the omni curve, but
the gain varies with elevation: $K$ peaks at $21.7$~dB near
$40^\circ$, then decreases to $17.5$~dB at $80^\circ$.

The ground and wall powers in \eqref{eq:ground_power} and
\eqref{eq:facade_power_weight} explain this peak.
The ground arrival moves away from boresight with offset
$2\bele$, so its antenna attenuation increases from $2.2$~dB
at $10^\circ$ to $34.7$~dB at $40^\circ$, suppressing the
dominant omni component in Fig.~\ref{fig:power_path}.
At higher elevations, wall-reflected signals arrive closer
to boresight and are therefore less attenuated by the antenna~\eqref {eq:wall_normal_projections}.
\endgroup

\subsection{Doppler Spread and Coherence Time}
\label{sec:res_compare}

\begin{figure}[!h]
    \centering
    \includegraphics[width=\linewidth]{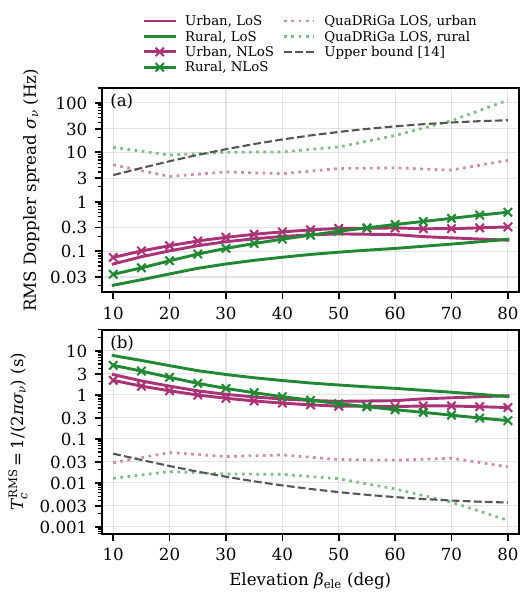}
    \caption{Non-direct (a) RMS Doppler spread and (b) coherence scale
    \eqref{eq:coherence_rms_scale}: LoS state (solid), NLoS state (dashed),
    QuaDRiGa LOS, and the~\cite{11534841} equal-power scatterer reference.}
    \label{fig:doppler_coherence_rms}
\end{figure}

\begingroup

In urban environments, the LoS-state Doppler spread increases with elevation and then gradually flattens. The NLoS-state spread follows a similar trend but remains higher. This difference arises because, in the LoS state, the ground reflection dominates the non-direct power and has a  concentrated residual Doppler given by \eqref{eq:ground_record}, whereas in the NLoS state, power is distributed over the broader Doppler contributions from wall reflections and edge diffraction.

In rural environments, the LoS- and NLoS-state Doppler spreads follow trends similar to those in urban environments, but their separation is larger. This is because the ground reflection carries about \(90\%\) of the rural LoS-state non-direct power, yielding a relatively concentrated Doppler distribution. 
\color{black}

In contrast, QuaDRiGa LOS gives Doppler spreads of similar magnitude to
our prior work~\cite{11534841}, which are wider than the proposed model. This agreement arises because the two models share similar underlying scatterer assumptions: both distribute scatterers in three-dimensional space, assign equal power to the scattered paths, and do not account for blockage. In contrast, the proposed model incorporates geometric constraints, path-dependent powers, and blockage, which jointly produce a narrower Doppler spread.
\endgroup

\subsection{Stationarity Interval}
\label{sec:res_stationarity}

\begingroup
Overall, $T_{\rm sta}$ tends to decrease with elevation
in both environments, with local increases near
$45^\circ$ in rural and $65^\circ$ in urban.

In urban scenarios, wall reflections mainly determine
$T_{\rm sta}$ at low and moderate elevations.
With further elevation increase, the wall contribution
weakens, consistent with the small rise in
$T_{\rm sta}$ from about $5.3$~s at $60^\circ$
to $6.4$~s at $65^\circ$.
At $\bele=65^\circ$--$85^\circ$, increasing ground
Doppler shift~\eqref{eq:ground_record} generally
shortens $T_{\rm sta}$ again, to about $5$~s.

In rural, ground reflection dominates the non-direct
power at all elevations. The local hump
reaches about $33$~s at $\bele=45^\circ$ due to the
slower variation of ground Doppler~\eqref{eq:ground_record}.
At higher elevations, faster ground Doppler drift
reduces $T_{\rm sta}$ to about $5.4$~s at $85^\circ$.
\endgroup

\begin{figure}[h]
    \centering
    \includegraphics[width=\linewidth]{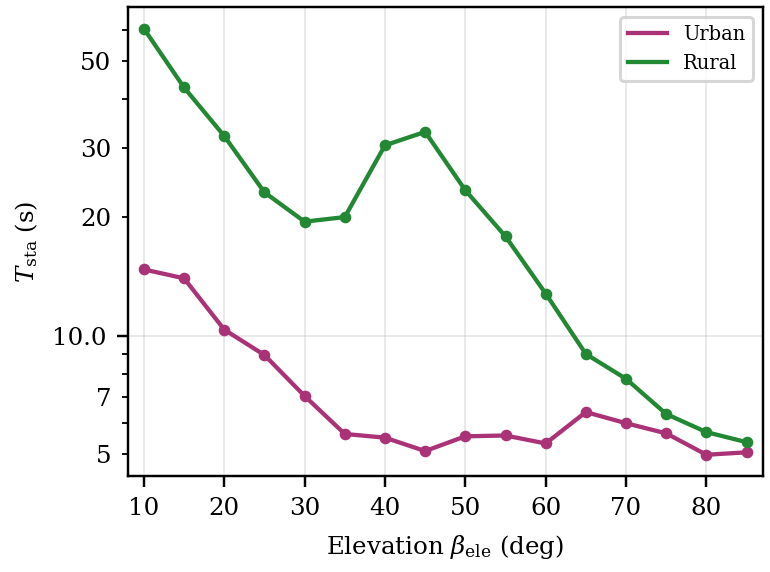}
    \caption{$T_{\rm sta}$ at $\bele=10^\circ$--$90^\circ$.}
    \label{fig:stationarity}
\end{figure}

\section{Conclusion}
This work developed a geometry-based stochastic LEO channel model
that maps a prescribed trajectory and ground scene
to an evolving CIR through wall and ground reflections and
roof-edge diffraction. Under the stated existence model,
the derived LoS/NLoS probabilities and state-conditioned
non-direct delay--Doppler densities yield elevation-dependent
delay and Doppler spreads, coherence time, and LoS-state $K$-factor.

The results show how geometry and existence shape NTN channel
statistics over a contact. Increasing LoS probability and
decreasing urban delay spread broadly follow TR~38.811, while
the urban delay-spread and increasing omnidirectional $K$-factor
trends broadly match 6G-NTN. Beamforming produces a mid-elevation
$K$-factor peak through ground rejection followed by greater
admission of wall power. Doppler spreads are smaller than those
from QuaDRiGa, partly due to the modeled existence conditions.
Stationarity times generally decrease with elevation but
temporarily increase as residual Doppler shift slows, providing guidance
for selecting local WSSUS intervals for TDL generation.

Together, these results offer a stochastic reference for
interpreting elevation-dependent channel behavior, complementing
site-specific ray-tracing models. Diffuse and vegetation scattering are omitted; rough-surface
scattering and link-level evaluation remain future work.

\section*{Acknowledgment}
The authors used Claude (Anthropic) to generate the simulation and analysis code used to produce the results in Section~\ref{sec:results}. The authors reviewed and tested the code and take full responsibility for the implementation and reported results.

\appendices
\begingroup
\setlength{\jot}{1pt}
\setlength{\abovedisplayskip}{4pt plus 1pt minus 1pt}
\setlength{\belowdisplayskip}{4pt plus 1pt minus 1pt}
\setlength{\abovedisplayshortskip}{0pt}
\setlength{\belowdisplayshortskip}{3pt plus 1pt minus 1pt}
\section{Reflection Power and Diffraction Coefficients}
\label{app:fresnel}
\label{app:utd}

\emph{Fresnel coefficients.}
For incidence angle {\footnotesize $\theta$} from the normal~\cite{ituP2040},
{\footnotesize
\begin{equation}
\begin{aligned}
&\Gamma_{\rm TE}(\theta,\widetilde\epsilon_r)
=\frac{\cos\theta-\sqrt{\widetilde\epsilon_r-\sin^2\theta}}
{\cos\theta+\sqrt{\widetilde\epsilon_r-\sin^2\theta}},\\
&\Gamma_{\rm TM}(\theta,\widetilde\epsilon_r)
=\frac{\widetilde\epsilon_r\cos\theta-\sqrt{\widetilde\epsilon_r-\sin^2\theta}}
{\widetilde\epsilon_r\cos\theta+\sqrt{\widetilde\epsilon_r-\sin^2\theta}}.
\end{aligned}
\label{eq:fresnel_coefficients}
\end{equation}}
Here {\footnotesize $\widetilde\epsilon_r=\epsilon_r'-j\sigma/(2\pi f_c\epsilon_0)$}, with relative permittivity {\footnotesize $\epsilon_r'$}, conductivity {\footnotesize $\sigma$},
vacuum permittivity {\footnotesize $\epsilon_0$}, and {\footnotesize $j^2=-1$}.

\emph{Wall power.}
For a large, smooth, planar wall, the image path in \eqref{eq:specular_direction}
has spreading factor {\footnotesize $(d_{s,i}^{\rm wall}+d_{g,i}^{\rm wall})^{-2}$}~\cite{tse_wireless_2005}.
Equal TE/TM weighting gives {\footnotesize $|\Gamma(\theta_{\rm s},\widetilde\epsilon_r)|^2
=\tfrac12[|\Gamma_{\rm TE}(\theta_{\rm s},\widetilde\epsilon_r)|^2
+|\Gamma_{\rm TM}(\theta_{\rm s},\widetilde\epsilon_r)|^2]$}.

Normalized by unobstructed LoS power, the Friis ratio is {\footnotesize $\frac{[c_0/(4\pi f_c(d_{s,i}^{\rm wall}+d_{g,i}^{\rm wall}))]^2}
{[c_0/(4\pi f_cR)]^2}
=\frac{(d_{s,i}^{\rm wall}+d_{g,i}^{\rm wall})^{-2}}{R^{-2}}$}~\cite{goldsmith_wireless_nodate}.

Equal transmit power and satellite gain give
{\footnotesize
\begin{equation}
W_i^{\rm wall}=|\Gamma(\theta_{\rm s},\widetilde\epsilon_r)|^2
\frac{(d_{s,i}^{\rm wall}+d_{g,i}^{\rm wall})^{-2}}{R^{-2}}
\frac{G(\vartheta_i)}{G(0)}.
\label{eq:wall_power_ratio_app}
\end{equation}}
With {\footnotesize $G(0)=1$}, \eqref{eq:wall_power_ratio_app} gives \eqref{eq:facade_power_weight},
while \eqref{eq:existence} weights it by visibility as {\footnotesize $\bm1_i^{\rm wall}W_i^{\rm wall}$}.

\emph{Roof-edge UTD coefficient.}
For the angles in Fig.~\ref{fig:edge_geometry}, use exterior angle {\footnotesize $n\pi$},
{\footnotesize $n=3/2$}, {\footnotesize $s_i(t)=\sqrt{1-[\hat{\mathbf e}_i^{\mathsf T}
\hat{\mathbf d}_{s,i}(t)]^2}>0$}, and {\footnotesize $L=d_{g,i}(t)s_i^2(t)$}.
Suppressing angular and material arguments, \eqref{eq:edge_power_weight}
uses~\cite{kouyoumjian_uniform_1974,ituP526}
{\footnotesize
\begin{equation}
\begin{aligned}
&D_i(t)={}-\frac{e^{-j\pi/4}}
{2n\,s_i(t)\sqrt{2\pi(2\pi f_c/c_0)}}\\
&\times\Bigg[
\sum_{\pm}\cot\!\left(\frac{\pi\pm(\varphi_2-\varphi_1)}{2n}\right)
F\!\left(\frac{2\pi f_c}{c_0}L a^\pm(\varphi_2-\varphi_1)\right)\\
&+\Gamma_0\cot\!\left(\frac{\pi-\varphi_2-\varphi_1}{2n}\right)
F\!\left(\frac{2\pi f_c}{c_0}L a^-(\varphi_2+\varphi_1)\right)\\
&+\Gamma_n\cot\!\left(\frac{\pi+\varphi_2+\varphi_1}{2n}\right)
F\!\left(\frac{2\pi f_c}{c_0}L a^+(\varphi_2+\varphi_1)\right)
\Bigg].
\end{aligned}
\label{eq:utd_coefficient}
\end{equation}}
Use {\footnotesize $F$} and {\footnotesize $a^\pm$} from~\cite[eqs.~(92), (97)--(99)]{ituP526},
radians, and combined cotangent--transition limits at ray boundaries~\cite[eq.~(102)]{ituP526}.

For TE, \eqref{eq:fresnel_coefficients} gives {\footnotesize $\Gamma_0=\Gamma_{\rm TE}(\arccos[s_i(t)|\sin\varphi_1|],
\widetilde\epsilon_r)$} and {\footnotesize $\Gamma_n=\Gamma_{\rm TE}(\arccos[s_i(t)|\cos\varphi_1|],
\widetilde\epsilon_r)$}.
For TM, replace {\footnotesize $\Gamma_{\rm TE}$} by {\footnotesize $\Gamma_{\rm TM}$} in \eqref{eq:utd_coefficient},
then average the two values of {\footnotesize $|D_i|^2$}.

\section{Derivation of Excess Delay and Residual Doppler}
\label{app:delay_doppler}
For a fixed scene and GT, suppress time, {\footnotesize $0<\bele<\pi/2$}.

\subsection{Wall Reflection}
\label{app:delay_doppler_wall}

\emph{1) Image geometry.}
From \eqref{eq:facade_frame},
{\footnotesize
\begin{equation}
\begin{aligned}
&\hat{\mathbf n}_i^{\mathsf T}\mathbf c_i
=\rho_i(-\sin\xi_i\cos\alpha_i+\cos\xi_i\sin\alpha_i) =\rho_i\sin(\alpha_i-\xi_i),\\
&\hat{\mathbf n}_i^{\mathsf T}\hat{\mathbf R}
=\cos\bele(-\sin\xi_i\cos\alpha_s+\cos\xi_i\sin\alpha_s)\\
&=\cos\bele\sin(\alpha_s-\xi_i).
\end{aligned}
\label{eq:wall_normal_projections}
\end{equation}}
{\footnotesize
\begin{equation}
\mathbf g_i'=[0,0,h_g]^{\mathsf T}
+2\rho_i\sin(\alpha_i-\xi_i)\hat{\mathbf n}_i.
\label{eq:wall_gt_image}
\end{equation}}
The image path in \eqref{eq:wall_gt_image} has length
{\footnotesize
\begin{equation}
d_{s,i}^{\rm wall}+d_{g,i}^{\rm wall}
=\left\|R\hat{\mathbf R}
-2\rho_i\sin(\alpha_i-\xi_i)\hat{\mathbf n}_i\right\|.
\label{eq:wall_image_length}
\end{equation}}

\emph{2) Path expansion.}
Triangle law: {\footnotesize $c^2=a^2+b^2-2ab\cos\theta$}; applying \eqref{eq:wall_normal_projections}
to \eqref{eq:wall_image_length} gives
{\footnotesize
\begin{equation}
\begin{aligned}
&(d_{s,i}^{\rm wall}+d_{g,i}^{\rm wall})^2
={}R^2+4\rho_i^2\sin^2(\alpha_i-\xi_i)\\
&-4R\rho_i\sin(\alpha_i-\xi_i)
\cos\bele\sin(\alpha_s-\xi_i).
\end{aligned}
\label{eq:wall_cosine_rule}
\end{equation}}
Taking the square root of \eqref{eq:wall_cosine_rule} gives
{\footnotesize
\begin{equation}
\begin{aligned}
&d_{s,i}^{\rm wall}+d_{g,i}^{\rm wall}
=R\Big[1
-\frac{4\rho_i}{R}\sin(\alpha_i-\xi_i)
\cos\bele\sin(\alpha_s-\xi_i)\\
&+\frac{4\rho_i^2}{R^2}\sin^2(\alpha_i-\xi_i)\Big]^{1/2}.
\end{aligned}
\label{eq:wall_sqrt_length}
\end{equation}}
Taylor expansion:
{\footnotesize
\begin{equation}
\sqrt{1+x}=1+\frac{x}{2}+O(x^2),\qquad |x|\ll1.
\label{eq:sqrt_first_order}
\end{equation}}
From \eqref{eq:wall_sqrt_length} and \eqref{eq:sqrt_first_order}, to first order: {\footnotesize $d_{s,i}^{\rm wall}+d_{g,i}^{\rm wall}
={}R-2\rho_i\sin(\alpha_i-\xi_i)
\cos\bele\sin(\alpha_s-\xi_i) +O(\rho_i^2/R).$}

Subtract {\footnotesize $R$} and divide by {\footnotesize $c_0$}:
{\footnotesize
\begin{equation}
\Delta\tau_i^{\rm wall}
\simeq-\frac{2\rho_i\sin(\alpha_i-\xi_i)
\cos\bele\sin(\alpha_s-\xi_i)}{c_0}.
\label{eq:image_delay_doppler}
\end{equation}}
The illuminated branch {\footnotesize $\sin(\alpha_i-\xi_i)\sin(\alpha_s-\xi_i)<0$} gives the positive form in
\eqref{eq:wall_delay_doppler}, subject to \eqref{eq:facade_rectangle}.

\emph{3) Doppler.}
With {\footnotesize $\rho_i,\alpha_i,\xi_i$} fixed, the product rule gives
{\footnotesize
\begin{equation}
\begin{aligned}
&\dot{\Delta\tau}_i^{\rm wall}
\simeq{}-\frac{2\rho_i\sin(\alpha_i-\xi_i)}{c_0}
\Big[-\dot\bele\sin\bele\sin(\alpha_s-\xi_i)\\
&+\dot\alpha_s\cos\bele\cos(\alpha_s-\xi_i)\Big]\\
&={}\left[-\frac{2\rho_i\sin(\alpha_i-\xi_i)
\cos\bele\sin(\alpha_s-\xi_i)}{c_0}\right]\\
&\times\left[-\dot\bele\frac{\sin\bele}{\cos\bele}
+\dot\alpha_s\frac{\cos(\alpha_s-\xi_i)}{\sin(\alpha_s-\xi_i)}\right]
\end{aligned}
\label{eq:wall_delay_derivative}
\end{equation}}

Multiply by {\footnotesize $-f_c$} as in \eqref{eq:residual_doppler} gives \eqref{eq:wall_delay_doppler}.

\subsection{Ground Reflection}
\label{app:delay_doppler_ground}
Mirroring the GT to {\footnotesize $[0,0,-h_g]^{\mathsf T}$} gives {\footnotesize $\|R\hat{\mathbf R}+[0,0,2h_g]^{\mathsf T}\|^2
=R^2+4h_g^2+4Rh_g\sin\bele,\quad\allowbreak \|R\hat{\mathbf R}+[0,0,2h_g]^{\mathsf T}\|
=R\sqrt{1+\frac{4h_g}{R}\sin\bele+\frac{4h_g^2}{R^2}} =R+2h_g\sin\bele+O(h_g^2/R).$}

The expansion uses \eqref{eq:sqrt_first_order} with {\footnotesize $x=4h_g\sin\bele/R+4h_g^2/R^2$}.

Subtract {\footnotesize $R$}, divide by {\footnotesize $c_0$}, and differentiate:
{\footnotesize
\begin{equation}
\begin{aligned}
&\Delta\tau^{\rm g}\simeq\frac{2h_g\sin\bele}{c_0},\quad
\dot{\Delta\tau}^{\rm g}\simeq\frac{2h_g\dot\bele\cos\bele}{c_0},\\
&\Delta\nu^{\rm g}=-f_c\dot{\Delta\tau}^{\rm g}
\simeq-\frac{2f_ch_g\dot\bele\cos\bele}{c_0}.
\end{aligned}
\label{eq:ground_delay_doppler_app}
\end{equation}}
This gives \eqref{eq:ground_record}.

\subsection{Roof-Edge Diffraction}
\label{app:delay_doppler_edge}

\emph{1) Edge geometry.}
The edge satisfies {\footnotesize $\hat{\mathbf n}_i^{\mathsf T}\mathbf p_i^{\rm edge}
=\rho_i\sin(\alpha_i-\xi_i)$} and {\footnotesize $[\mathbf p_i^{\rm edge}]_z=H_i$}.

Substitute {\footnotesize $\hat{\mathbf d}_{s,i}^{\rm edge}\simeq-\hat{\mathbf R}$} and {\footnotesize $\hat{\mathbf d}_{g,i}^{\rm edge}
=([0,0,h_g]^{\mathsf T}-\mathbf p_i^{\rm edge})/d_{g,i}^{\rm edge}$} into \eqref{eq:29}:
{\footnotesize
\begin{equation}
-\frac{\hat{\mathbf e}_i^{\mathsf T}\mathbf p_i^{\rm edge}}
{d_{g,i}^{\rm edge}}
\simeq-\hat{\mathbf e}_i^{\mathsf T}\hat{\mathbf R},\quad \hat{\mathbf e}_i^{\mathsf T}\mathbf p_i^{\rm edge}
\simeq d_{g,i}^{\rm edge}\cos\bele\cos(\alpha_s-\xi_i).
\label{eq:edge_along_component}
\end{equation}}
Summing squared components with \eqref{eq:edge_along_component} gives {\footnotesize $(d_{g,i}^{\rm edge})^2 \simeq\rho_i^2\sin^2(\alpha_i-\xi_i)
+(d_{g,i}^{\rm edge})^2\cos^2\bele\cos^2(\alpha_s-\xi_i) +(H_i-h_g)^2,\quad\allowbreak (d_{g,i}^{\rm edge})^2
[1-\cos^2\bele\cos^2(\alpha_s-\xi_i)] \simeq\rho_i^2\sin^2(\alpha_i-\xi_i)+(H_i-h_g)^2.$}

Divide by the bracket and take the positive root: {\footnotesize $d_{g,i}^{\rm edge}\simeq
\sqrt{\frac{\rho_i^2\sin^2(\alpha_i-\xi_i)+(H_i-h_g)^2}
{1-\cos^2\bele\cos^2(\alpha_s-\xi_i)}}.$}

Enforce the finite-edge condition \eqref{eq:edge_cond}.

\emph{2) Path expansion.}
Triangle law gives {\footnotesize $(d_{s,i}^{\rm edge})^2
=R^2+(d_{g,i}^{\rm edge})^2-2Rd_{g,i}^{\rm edge}\cos\vartheta_i,\quad\allowbreak d_{s,i}^{\rm edge}
=R\sqrt{1-\frac{2d_{g,i}^{\rm edge}}{R}\cos\vartheta_i
+\frac{(d_{g,i}^{\rm edge})^2}{R^2}} =R-d_{g,i}^{\rm edge}\cos\vartheta_i
+O((d_{g,i}^{\rm edge})^2/R).$}

The expansion uses \eqref{eq:sqrt_first_order} with {\footnotesize $x=-2d_{g,i}^{\rm edge}\cos\vartheta_i/R+(d_{g,i}^{\rm edge}/R)^2$}.

Add {\footnotesize $d_{g,i}^{\rm edge}$}, subtract {\footnotesize $R$}, and divide by {\footnotesize $c_0$}:
{\footnotesize
\begin{equation}
\Delta\tau_i^{\rm edge}
\simeq\frac{d_{g,i}^{\rm edge}(1-\cos\vartheta_i)}{c_0}.
\label{eq:edge_delay_app}
\end{equation}}
The LoS projection is
{\footnotesize
\begin{equation}
d_{g,i}^{\rm edge}\cos\vartheta_i
=\hat{\mathbf R}^{\mathsf T}
(\mathbf p_i^{\rm edge}-[0,0,h_g]^{\mathsf T}).
\label{eq:edge_los_projection}
\end{equation}}

\emph{3) Doppler.}
Differentiating {\footnotesize $(d_{g,i}^{\rm edge})^2
=\|\mathbf p_i^{\rm edge}-[0,0,h_g]^{\mathsf T}\|^2$} gives {\footnotesize $2d_{g,i}^{\rm edge}\dot d_{g,i}^{\rm edge}
=2(\mathbf p_i^{\rm edge}-[0,0,h_g]^{\mathsf T})^{\mathsf T}
\dot{\mathbf p}_i^{\rm edge},\quad\allowbreak \dot d_{g,i}^{\rm edge}
=\frac{(\mathbf p_i^{\rm edge}-[0,0,h_g]^{\mathsf T})^{\mathsf T}}
{d_{g,i}^{\rm edge}}\dot{\mathbf p}_i^{\rm edge}.$}

Substituting \eqref{eq:edge_los_projection} into \eqref{eq:edge_delay_app}
and applying the product rule gives
{\footnotesize
\begin{equation}
\begin{aligned}
&c_0\dot{\Delta\tau}_i^{\rm edge}
\simeq{}\dot d_{g,i}^{\rm edge}
-\dot{\hat{\mathbf R}}{}^{\mathsf T}
(\mathbf p_i^{\rm edge}-[0,0,h_g]^{\mathsf T})
-\hat{\mathbf R}^{\mathsf T}\dot{\mathbf p}_i^{\rm edge}\\
&={}-\dot{\hat{\mathbf R}}{}^{\mathsf T}
(\mathbf p_i^{\rm edge}-[0,0,h_g]^{\mathsf T})+\left(\frac{\mathbf p_i^{\rm edge}-[0,0,h_g]^{\mathsf T}}
{d_{g,i}^{\rm edge}}-\hat{\mathbf R}\right)^{\mathsf T}
\dot{\mathbf p}_i^{\rm edge}.
\end{aligned}
\label{eq:edge_delay_derivative_app}
\end{equation}}
By \eqref{eq:edge_along_component}, {\footnotesize $\hat{\mathbf e}_i^{\mathsf T}
\left(\frac{\mathbf p_i^{\rm edge}-[0,0,h_g]^{\mathsf T}}
{d_{g,i}^{\rm edge}}-\hat{\mathbf R}\right)
\simeq0,$} multiply by {\footnotesize $-f_c/c_0$}:
{\footnotesize
\begin{equation}
\Delta\nu_i^{\rm edge}\simeq\frac{f_c}{c_0}
\dot{\hat{\mathbf R}}{}^{\mathsf T}
(\mathbf p_i^{\rm edge}-[0,0,h_g]^{\mathsf T}).
\label{eq:edge_doppler_app}
\end{equation}}
This gives \eqref{eq:edge_delay_doppler}.

\begin{samepage}
\section{Mean Visibility Indicators}
\label{app:visibility}

\emph{Wall reflection.}
Since building heights are i.i.d\\ {\small $\mathbb E[\bm1_i^{\rm wall}]
={}\Pr\{\text{specular within wall width}\} \times\Pr\{H_i\geq[\mathbf p_i^{\rm wall}]_z\} \times\Pr\{\text{no {\footnotesize $j\ne i$} blocks the GT segment}
\mid\mathbf p_i^{\rm wall}\}.$}
\par
\end{samepage}

\emph{First term}, wall footprints of length {\footnotesize $\ell_{\rm f}$}
at angle {\footnotesize $\xi_j\sim\mathcal U[0,\pi)$} to the horizontal path yield
{\footnotesize $\lambda_b\ell_{\rm f}|\sin\xi_j|$}
crossings per unit horizontal path length on average.
Averaging gives~\cite{bai_blockage_2014}:
{\footnotesize
\begin{equation}
\frac{1}{\Delta\rho}
\mathbb E_{\xi_j}\!\left[
\lambda_b\,\Delta\rho\,\ell_{\rm f}|\sin\xi_j|
\right]
=\frac{\lambda_b\ell_{\rm f}}{2\pi}
\int_0^{2\pi}|\sin\xi_j|\,d\xi_j
=\frac{2\lambda_b\ell_{\rm f}}{\pi}.
\label{eq:blocker_crossing_rate}
\end{equation}}

\emph{Second term}, is  {\footnotesize $\Pr\{H_i\geq[\mathbf p_i^{\rm wall}]_z\}
=e^{-[\mathbf p_i^{\rm wall}]_z^2/(2\gamma^2)}$} from \eqref{eq:itu_height_ccdf}.

 \emph{Third term}, 
 Use {\footnotesize $z=h_g+\rho\tan\bele$} and {\footnotesize $d\rho=dz/\tan\bele$}, the Poisson zero-count probability gives
{\footnotesize
\begin{equation}
\begin{aligned}
&\Pr\{\text{no {\footnotesize $j\ne i$} blocks the GT segment}
\mid\mathbf p_i^{\rm wall}\}\\
&\simeq\exp\!\left[-\frac{2\lambda_b\ell_{\rm f}}{\pi}
\int_0^{\|[\mathbf p_i^{\rm wall}]_{xy}\|}
e^{-(h_g+\rho\tan\bele)^2/(2\gamma^2)}\,d\rho\right]\\
&=\exp\!\left[-\frac{2\lambda_b\ell_{\rm f}}{\pi\tan\bele}
\int_{h_g}^{[\mathbf p_i^{\rm wall}]_z}e^{-z^2/(2\gamma^2)}\,dz\right],
\end{aligned}
\label{eq:blocker_count}
\end{equation}}
where
{\footnotesize
\begin{equation}
\int_{h_g}^{[\mathbf p_i^{\rm wall}]_z}e^{-z^2/(2\gamma^2)}\,dz
=\gamma\sqrt{\frac{\pi}{2}}
\left[\operatorname{erfc}\frac{h_g}{\sqrt2\gamma}
-\operatorname{erfc}\frac{[\mathbf p_i^{\rm wall}]_z}{\sqrt2\gamma}\right].
\label{eq:height_integral_erfc}
\end{equation}}
Note that {\footnotesize $\frac{d}{dz}\operatorname{erfc}\!\left(\frac{z}{\sqrt2\gamma}\right)
=-\frac{\sqrt{2/\pi}}{\gamma}e^{-z^2/(2\gamma^2)}.$}

Thus, with {\footnotesize $2\sqrt{\pi/2}/\pi=\sqrt{2/\pi}$}:
{\footnotesize
\begin{equation}
\begin{aligned}
&\mathbb E[\bm1_i^{\rm wall}]
\simeq e^{-[\mathbf p_i^{\rm wall}]_z^2/(2\gamma^2)}\\
&\times\exp\!\Biggl[-\frac{\sqrt{2/\pi}\lambda_b\ell_{\rm f}\gamma}{\tan\bele}
\left(\operatorname{erfc}\frac{h_g}{\sqrt2\gamma}
-\operatorname{erfc}\frac{[\mathbf p_i^{\rm wall}]_z}{\sqrt2\gamma}\right)\Biggr].
\end{aligned}
\label{eq:wall_indicator_closed_app}
\end{equation}}
This gives \eqref{eq:visibility_wall}.

\emph{Direct and ground paths.}
At horizontal distance $\rho$ from the GT along the projected direct ray,
the ray height above the ground is $h_g+\rho\tan\bele$.
The direct path is visible if every wall height $H_j<h_g+\rho\tan\bele$ at its crossing distance
$\rho$. Replacing the upper
integration limit in \eqref{eq:blocker_count} by $R_{\rm city}$ gives
{\footnotesize
\begin{equation}
\begin{aligned}
\mathbb E[\bm1_0]
&\simeq \exp\!\left[
-\frac{2\lambda_b\ell_{\rm f}}{\pi}
\int_0^{R_{\rm city}}
e^{-(h_g+\rho\tan\bele)^2/(2\gamma^2)}\,d\rho
\right]\\
&= \exp\!\left[
-\frac{2\lambda_b\ell_{\rm f}}{\pi\tan\bele}
\int_{h_g}^{h_g+R_{\rm city}\tan\bele}
e^{-z^2/(2\gamma^2)}\,dz
\right]\\
&= \exp\!\Biggl[
-\frac{\sqrt{2/\pi}\lambda_b\ell_{\rm f}\gamma}{\tan\bele}
\Biggl(
\operatorname{erfc}\!\frac{h_g}{\sqrt2\gamma}
-\operatorname{erfc}\!\frac{h_g+R_{\rm city}\tan\bele}{\sqrt2\gamma}
\Biggr)
\Biggr].
\end{aligned}
\end{equation}}
The second equality uses
{\footnotesize $z=h_g+\rho\tan\bele$}, and the third follows
from \eqref{eq:height_integral_erfc}.
When {\footnotesize $h_g+R_{\rm city}\tan\bele\gg\gamma$},
the second erfc term is negligible, yielding
{\footnotesize
\begin{equation}
\mathbb E[\bm1_0]\simeq
\exp\!\left[
-\frac{\sqrt{2/\pi}\lambda_b\ell_{\rm f}\gamma}{\tan\bele}
\operatorname{erfc}\!\frac{h_g}{\sqrt2\gamma}
\right].
\label{eq:direct_indicator_closed_app}
\end{equation}}
This gives \eqref{eq:blockage_B}.

\emph{Roof-edge diffraction.}
At fixed roof height, \eqref{eq:existence} gives
{\footnotesize
\begin{equation}
\begin{aligned}
&\mathbb E[\bm1_i^{\rm edge}\mid H_i] =\Pr\{\text{edge geometry}\}\\
&\times\Pr\{\text{no {\footnotesize $j\ne i$} blocks the GT segment}
\mid\mathbf p_i^{\rm edge}\}.
\end{aligned}
\label{eq:edge_indicator_mean_app}
\end{equation}}
Using segment height {\footnotesize $h_g+\rho(H_i-h_g)/\|[\mathbf p_i^{\rm edge}]_{xy}\|$} to \eqref{eq:blocker_count} gives
{\footnotesize
\begin{equation}
\begin{aligned}
&\Pr\{\text{no {\footnotesize $j\ne i$} blocks the GT segment}
\mid\mathbf p_i^{\rm edge}\}\\
&\simeq\exp\!\Biggl[-\frac{2\lambda_b\ell_{\rm f}}{\pi}
\int_0^{\|[\mathbf p_i^{\rm edge}]_{xy}\|}\exp\!\left(-\frac{\left[h_g+\rho\frac{H_i-h_g}
{\|[\mathbf p_i^{\rm edge}]_{xy}\|}\right]^2}{2\gamma^2}\right)d\rho\Biggr]\\
&=\exp\!\left[-\frac{2\lambda_b\ell_{\rm f}\|[\mathbf p_i^{\rm edge}]_{xy}\|}
{\pi(H_i-h_g)}\int_{h_g}^{H_i}e^{-z^2/(2\gamma^2)}\,dz\right].
\end{aligned}
\label{eq:edge_blocker_integral}
\end{equation}}
For {\footnotesize $H_i\ne h_g$}, substitute {\footnotesize $z=h_g+\rho(H_i-h_g)/\|[\mathbf p_i^{\rm edge}]_{xy}\|$} and {\footnotesize $d\rho=\|[\mathbf p_i^{\rm edge}]_{xy}\|\,dz/(H_i-h_g)$}.

Applying \eqref{eq:height_integral_erfc} with upper endpoint {\footnotesize $H_i$}
in \eqref{eq:edge_indicator_mean_app} gives
{\footnotesize
\begin{equation}
\begin{aligned}
&\mathbb E[\bm1_i^{\rm edge}\mid H_i]
\simeq\Pr\{\text{edge geometry}\}\\
&\times\exp\!\Biggl[-\frac{\sqrt{2/\pi}\lambda_b\ell_{\rm f}\gamma
\|[\mathbf p_i^{\rm edge}]_{xy}\|}{H_i-h_g}
\left(\operatorname{erfc}\frac{h_g}{\sqrt2\gamma}
-\operatorname{erfc}\frac{H_i}{\sqrt2\gamma}\right)\Biggr].
\end{aligned}
\label{eq:edge_indicator_closed_app}
\end{equation}}
The probability $\Pr\{\text{edge geometry\}}$ is evaluated
numerically with
\eqref{eq:edge_cond} and \eqref{eq:29}.



\endgroup

\bibliographystyle{IEEEtran}
\bibliography{reference}

\end{document}